\documentclass[twocolumn,prb,floatfix,showpacs, bibnotes]{revtex4}
\usepackage{graphicx}
\usepackage{amsmath, amssymb, bm}
\usepackage{pstricks}
\usepackage{multirow}

\newcommand{\one}{\ensuremath{^{(1)}}}
\newcommand{\zer}{\ensuremath{^{(0)}}}

\newcommand{\bra}[1]{\ensuremath{\langle#1|}}
\newcommand{\ket}[1]{\ensuremath{{|#1\rangle}}}
\newcommand{\braket}[2]{\ensuremath{\langle#1|#2\rangle}}
\renewcommand{\vec}[1]{{\ensuremath{\bm{\mathrm{#1}}}}}

\begin{document}
\title{Spin and orbital magnetic response in metals: susceptibility and NMR shifts}
\author{Mayeul d'Avezac$^1$}
\author{Nicola Marzari$^2$}
\author{Francesco Mauri$^1$}
\affiliation{$^1$Institut de Min\'eralogie et Physique des Milieux Condens\'e, 
             case 115, 4 place Jussieu, 75252, Paris cedex 05, France}
\affiliation{$^2$Department of Materials Science and Engineering, 
             Massachusetts Institute of Technology, Cambridge, Massachusetts 02139-4307}
\date{\today}

\begin{abstract}
  A DFT-based method is presented which allows the computation of
  all-electron NMR shifts of metallic compounds with periodic boundary
  conditions. NMR shifts in metals measure two competing
  physical phenomena. Electrons interact with the applied magnetic
  field (i) as magnetic dipoles (or spins), resulting in the Knight
  shift, (ii) as moving electric charges, resulting in the chemical
  (or  orbital) shift. The latter is treated through an extension to
  metals of the Gauge Invariant Projector Augment Wave(GIPAW)
  developed for insulators. The former is modeled as the hyperfine
  interaction between the electronic spin polarization and the nuclear
  dipoles. NMR shifts are obtained with respect to the computed shieldings
  of reference compounds, yielding fully ab-initio quantities which
  are directly comparable to experiment. The method is validated by
  comparing the magnetic susceptibility of interacting and
  non-interacting homogeneous gas with known analytical results, and
  by comparing the computed NMR shifts of simple metals with
  experiment. 
\end{abstract}

\maketitle

\section{Introduction}

Nuclear Magnetic Resonance (NMR) is a widely used and powerful
technique for structural determination, both in chemistry and in
solid-state physics\cite{grant}. It also yields valuable information
on the electronic structure of solids. For instance, NMR was
instrumental in determining the $d_{x^2-y^2}$ pairing of
high-temperature superconductors\cite{pines:1997}. Empirical rules
have been determined  which relate NMR quantities to physical and
chemical properties. Unfortunately, such rules can become inaccurate
when subtle quantum effects are involved. In this work, we provide a
method for computing NMR shifts from first-principles in metallic
systems with periodic boundary conditions. 

Recent advances have made possible the computation of NMR shifts in
molecules\cite{keith:1992} and insulating solids with periodic
boundary conditions\cite{pickard:2001, sebastiani:2001}, leading to a
better interpretation of experimental data in systems as diverse as
zeolite\cite{profeta:2003} or vitreous Boron
oxides\cite{umari:2005}. 

At present, to the best of the authors' knowledge, there is no complete
\textit{ab-initio} theory of NMR shifts in \textit{metallic} systems. Indeed,
NMR shifts in metals result from two different physical phenomenon. The
electronic structure can react to the external magnetic field (i) as a
distribution of magnetic spins, giving rise to the Knight shift, (ii) as a
distribution of electronic charges, with the NMR orbital shift as a result. In
most metallic systems, the NMR shift is dominated by the Knight shift
contribution, sometimes by as much as two orders of magnitude. As such, it has
been the subject of many theoretical studies\cite{tripathi:1981,pavarini:2001}.
On the other hand, the development of methods capable of computing orbital
shifts in metallic compounds has been lagging behind. Yet, experiments do not
distinguish between the shifts arising from these two phenomena. Furthermore,
experimental shifts are given with respect to some insulating
reference-compound. As such, theoretical calculations must include both orbital
and Knight shifts in the material of interest and a reference-compound before
being compared to experiment. The Knight shift is related to the density of
$s$-states at the Fermi level. As such, there are a number
of systems for which the Knight and orbital contributions to NMR shifts and to the 
magnetic susceptibility are of similar magnitude. These systems include
semi-metals such as graphene, graphite\cite{lauginie:1988, hyroyama:1988},
intercalated graphite\cite{lauginie:1993, kobayashi:1988}, and
nanotubes\cite{ajiki:1995}, metals with strong $d$-character such as Platinum
catalysts\cite{stokes:1982-i, stokes:1982-iii}, or organic compounds adsorbed
upon metallic catalysts\cite{makowka:1985, vuissoz:1999}. 

The aim of the method presented here, is to provide a unified
first-principles framework to compute both orbital and Knight shifts
in metallic systems with periodic boundary conditions. The setting
for the method is density functional theory (DFT) as implemented in
plane-wave, pseudo-potential codes. The projector-augmented wave (PAW)
approach\cite{blochl:1994} allows us to  obtain accurate results from
pseudo-potential quantities. The problem of gauge invariance in
periodic pseudo-potential systems is treated using the gauge-invariant projector
augmented wave (GIPAW) approach of Ref.~\onlinecite{pickard:2001}. Our method
is entirely self-contained in the sense that we can compute the NMR
shielding of both metallic compounds of interest and the NMR shielding
of reference compounds. As such, the resulting NMR shifts are
directly comparable to experimental results.

The paper is organized as follows. First, we go over the physics
involved in computing NMR shifts. Secondly, we briefly review the
so-called ``smearing technique\cite{gironcoli:1995}'' which allows an
accurate and efficient treatment of the Fermi surface. We then
detail the computation of the orbital shift in sec.~\ref{sec:orb}, and
of the Knight shift in sec.~\ref{sec:knight}. In
sec.~\ref{sec:practical}, we discuss practical issues dealing with the
actual implementation of the method. The next section is devoted to
the study of limit-systems and numerical tests. Finally, the last
section presents results obtained on simple metals.

\section{NMR shifts in metals}

A uniform external magnetic field $\vec{B}$ applied to a metallic material
generates two different electronic behaviors: (i) a so-called \emph{orbital}
response where electrons react to the field as \textit{moving} charges, (ii) a
\textit{spin} response where electrons react as \textit{spinning} charges. \\

In the following and throughout the paper, we use the symmetric gauge
$\vec{A}(\vec{r}) =-\vec{B}\wedge\vec{r}$, with $\vec{A}$ the gauge,
$\vec{B}$ the magnetic field, and $\vec{r}$ the position in real-space.

The applied magnetic field induces an orbital current $\vec{j}_o(\vec{r}')$. It
can be obtained as the expectation value of the current operator
$\vec{J}(\vec{r}')$,
\begin{gather}
  \label{eq:joperator}
  \vec{J}(\vec{r}')= \vec{J}^p(\vec{r}')
                     -\frac{\vec{B}\wedge\vec{r}'}{2c}\ket{\vec{r}'}\bra{\vec{r}'}\\
  \vec{J}^p(\vec{r}')= -\frac{1}{2}\left(\vec{p}\ket{\vec{r}'}\bra{\vec{r}'} +
                                       \ket{\vec{r}'}\bra{\vec{r}'}\vec{p}\right)
\end{gather}
$c$ is the speed of light.  The first term on the right-hand-side of Eq.~\ref{eq:joperator}
is the paramagnetic current operator. The second term is the diamagnetic
current operator as expressed within the symmetric gauge. Note that at zero
field, the expectation value of $\vec{J}(\vec{r}')$ is null.

The orbital current induces in turn an inhomogeneous field
$\vec{B}_i(\vec{r}')$, which can be obtained from classical magnetostatics,
\begin{equation}
  \vec{B}_o(\vec{r}') = \frac{1}{c}\int\rm{d}\vec{r}\ \label{eq:field}
                           \vec{j}_o(\vec{r})\wedge\frac{\vec{r}'-\vec{r}}{|\vec{r}'-\vec{r}|^3}.
\end{equation}

We will describe our approach to the calculation of an all-electron induced
orbital current $\vec{j}_o$ using pseudopotentials in section
\ref{sec:orb}. The method is an extension to metals of the scheme proposed in
Ref.~\onlinecite{pickard:2001}. \\

The spin response results from a spin polarization of the electronic cloud by the
external magnetic field. To compute the resulting net electronic magnetization
$\vec{m}(\vec{r}')$, we make a co-linearity hypothesis, whereby $\vec{m}(\vec{r}')$ is supposed
parallel to the applied magnetic field. This hypothesis is used routinely in
hyperfine parameter and Knight shift calculations\cite{blochl:1993, pacchioni:2000,
pavarini:2001}. Hence, $\vec{m}(\vec{r}')$ can be obtained as,
\begin{equation} 
  \vec{m}(\vec{r}') = \frac{1}{2c}\left[ \rho_\uparrow(\vec{r}')
                      -\rho_\downarrow(\vec{r}')\right] \frac{\vec{B}}{B},
\end{equation}
where $\rho_\uparrow(\vec{r}')$ and $\rho_\downarrow(\vec{r}')$ are the up and
down spin densities.  The electronic magnetization induces a magnetic field
$\vec{B}_s$ which can be obtained form classical magnetostatics
\begin{multline}
  \label{eq:dipolefield}
  \vec{B}_s(\vec{r}') = \frac{8\pi}{3}\int\rm{d}\vec{r}\ \delta(\vec{r}'-\vec{r})\vec{m}(\vec{r}) \\
                        + \int\rm{d}\vec{r}\left\{
                          \frac{3\vec{m}(\vec{r})\cdot(\vec{r}'-\vec{r})}{|\vec{r}'-\vec{r}|^5}
                          - \frac{\vec{m}(\vec{r})}{|\vec{r}'-\vec{r}|^3}\right\}
\end{multline}
$\vec{B}_s(\vec{r}')$ is composed of two terms: (i) an on-site term, called the
Fermi contact (first term in Eq.~\ref{eq:dipolefield}) representing the dipole-field
at $\vec{r}'$, (ii) a long-distance dipolar term resulting from the full
magnetic dipole distribution $\vec{m}(\vec{r}')$.

A method to compute the electronic magnetization $\vec{m}(\vec{r})$ to first
order in $B$ is given in section \ref{sec:knight}. \\

For field strengths in the range typical to NMR, the orbital and spin responses
can be computed separately and to first order in $B$. The resulting linear
relationships between the induced first-order fields $\vec{B}\one_o(\vec{r}')$ and
$\vec{B}\one_s(\vec{r}')$,  and the external magnetic field $\vec{B}$ define
the orbital and spin shielding tensors, respectively
$\tensor{\sigma}_o(\vec{r}')$ and $\tensor{\sigma}_s(\vec{r}')$.
\begin{equation}
  \vec{B}\one_o(\vec{r}') = -\tensor{\sigma}_o(\vec{r}')\cdot\vec{B} \quad; \quad
  \vec{B}\one_s(\vec{r}') = -\tensor{\sigma}_s(\vec{r}')\cdot\vec{B} \label{eq:b}
\end{equation}
The isotropic NMR shielding $\sigma(\vec{R})$ of the nucleus at position
$\vec{R}$ is given by the trace $\sigma(\vec{R}) =
\rm{Tr}[\tensor{\sigma}_o(\vec{R})+\tensor{\sigma}_s(\vec{R})]/3$. The isotropic
NMR shift $\delta$, i.~e.\ the experimental observable, is obtained with
respect to the isotropic shielding $\sigma_{ref}$ of a so-called \emph{zero-shift compound},
with $\delta(\vec{R}) = -(\sigma(\vec{R}) - \sigma_{ref})$.\\ 

\subsection{Pseudo-potential System}
Within a pseudo-potential system, one must define the Hamiltonian and operators
with care. Following the projector augmented wave method\cite{blochl:1994}
(PAW), and the gauge including projector augmented wave
method\cite{pickard:2001} (GIPAW), the spin-Hamiltonian of a system with a
homogeneous magnetic field becomes
\begin{multline}
  \bar{\mathcal{H}}_\sigma = \frac{1}{2}\left( \vec{p} + \frac{\vec{B}\wedge\vec{r}}{2c} \right)^2
                + \mathcal{V}_{scf}^\sigma\\
                + \sum_\vec{R} e^{\frac{\imath}{2c}\vec{r}\cdot\vec{R}\wedge\vec{B}}
                  \mathcal{V}^{nl}_\vec{R} e^{-\frac{\imath}{2c}\vec{r}\cdot\vec{R}\wedge\vec{B}}
                + \rm{sgn}(\sigma)\frac{g_eB}{4c}.
\end{multline}
$\sigma$ indicates the spin-channel. $\vec{p}$ is the kinetic energy operator,
$\mathcal{V}_{scf}^\sigma$ the magnetic-field-dependent self-consistent potential, and
$\mathcal{V}_\vec{R}^{nl}$ the non-local potential at position
$\vec{R}$.  $\rm{sgn}(\sigma)$ returns $\pm 1$ depending on the spin
channel.  $g_e=2.0023193$ is the gyromagnetic ratio of the free
electron. The bar above quantities such as $\vec{\bar{H}}_\sigma$
indicates pseudo-potential reconstructed operators. The above can be
expanded to first order in $B$ as,
\begin{gather}
  \bar{\mathcal{H}}_\sigma = \bar{\mathcal{H}}\zer +
                             \bar{\mathcal{H}}\one_{s\sigma} + \bar{\mathcal{H}}\one_o 
                             + O(B^2),\label{eq:ham}\\ 
  \bar{\mathcal{H}}\zer = \frac{1}{2}\vec{p}^2
                          + \mathcal{V}_{scf}\zer
                          + \sum_\vec{R}\mathcal{V}^{nl}_\vec{R},\label{eq:hamzer}\\
  \bar{\mathcal{H}}_o\one = \frac{1}{2c}\left( \vec{L} + \label{eq:hamorb}
                            \sum_\vec{R}\vec{R}\wedge\vec{v}^{nl}_\vec{R} \right)\cdot\vec{B},\\
  \bar{\mathcal{H}}_{s\sigma}\one =
  \rm{sgn}(\sigma)B\left(\frac{g_e}{4c} +
  \mathcal{V}_{scf}\one(\vec{r})\right).\label{eq:hamspin}
\end{gather}
Note that we are interested in systems which are spin-degenerate at zero field,
hence $\bar{\mathcal{H}}\zer$ is defined independent of spin and does not carry a
spin index. $\vec{v}^{nl}_\vec{R}$ is a reconstruction
term\cite{pickard:2001}  defined as 
\begin{equation}
  \vec{v}^{nl}_\vec{R} = \frac{1}{\imath}\left[ \vec{r},\mathcal{V}^{nl}_\vec{R} \right].
\end{equation}
Square brackets indicate a commutator. $\vec{L}=\vec{r}\wedge\vec{p}$ is the
orbital momentum operator.  In the expansion above first order terms are
separated into  a spin dependent pertubation $\mathcal{H}_{s\sigma}\one$ and an
orbital dependent term $\mathcal{H}_o\one$. The former is given within the
colinear hypothesis discussed in the previous section. $\mathcal{H}_{s\sigma}\one$ is the 
only spin-dependent term in the expansion to first order in $B$ of
$\bar{\mathcal{H}}_\sigma$. $\mathcal{V}\one_{scf}$ is the linear part of the
self-consistent potential with respect to $\vec{B}$. It is obtained from the
functional derivative of $\mathcal{V}_{scf}^\sigma$ with respect
to the first-order electronic magnetization $m\one$ at zero field,
\begin{equation}
  \label{eq:potscf}
  \mathcal{V}_{scf}\one(\vec{r}') = \int\rm{d}\vec{r}\ \frac{1}{2}
    \frac{\partial\sum_\sigma\mathcal{V}_{scf}^\sigma(\vec{r}')}{\partial
    m(\vec{r})}m\one(\vec{r}),
\end{equation}

In order to obtain all-electron NMR shifts, we should also reconstruct the
current operator and the electronic magnetization. The former can be expressed to first order
as in Ref.~\onlinecite{pickard:2001},
\begin{equation} 
  \bar{\vec{J}}(\vec{r}') = \label{eq:curps}
          \bar{\vec{J}}\zer(\vec{r}') + \bar{\vec{J}}\one(\vec{r}') + O(B^2),
\end{equation}
with
\begin{equation}
  \bar{\vec{J}}\zer(\vec{r}') = \vec{J}^p(\vec{r}') + \sum_\vec{R}\Delta\vec{J}^p_\vec{R}(\vec{r}'),
\end{equation}
and
\begin{multline}
  \bar{\vec{J}}\one(\vec{r}') = \frac{\vec{B}\wedge\vec{r}'}{2c}\ket{\vec{r}'}\bra{\vec{r}'} \\+ 
      \sum_\vec{R}\left[\Delta\vec{J}^d_\vec{R}(\vec{r}') 
      + \frac{1}{2\imath c}[\vec{R}\wedge\vec{R}\cdot\vec{r}, 
                            \Delta\vec{J}^p_\vec{R}(\vec{r}')]\right].
\end{multline}
The paramagnetic reconstruction operator $\Delta\vec{J}_\vec{R}^p(\vec{r}')$
and diamagnetic reconstruction operator $\Delta\vec{J}^d_\vec{R}(\vec{r}')$ are
defined as follows,
{\small
\begin{multline}
  \Delta\vec{J}^p_\vec{R}(\vec{r}') = \sum_{n,m} \ket{\tilde{p}_{\vec{R},n}}\bigl[
       \bra{\Phi_{\vec{R},n}}\vec{J}^p(\vec{r}')\ket{\Phi_{\vec{R},m}} \\ -
       \bra{\tilde{\Phi}_{\vec{R},n}}\vec{J}^p(\vec{r}')\ket{\tilde{\Phi}_{\vec{R},m}} \bigr]
       \bra{\tilde{p}_{\vec{R},m}},
\end{multline}}
and
{\small
\begin{multline}
  \Delta\vec{J}^d_\vec{R}(\vec{r}') = -\frac{\vec{B}\wedge(\vec{r}'-\vec{R})}{2c}\sum_{n,m} 
    \ket{\tilde{p}_{\vec{R},n}}\bigl[
       \braket{\Phi_{\vec{R},n}}{\vec{r}'}\braket{\vec{r'}}{\Phi_{\vec{R},m}} \\ -
       \braket{\tilde{\Phi}_{\vec{R},n}}{\vec{r}'}\braket{\vec{r}'}{\tilde{\Phi}_{\vec{R},m}} \bigr]
       \bra{\tilde{p}_{\vec{R},m}}\label{eq:diam}.
\end{multline}}
The projector functions $\ket{\tilde{p}_{\vec{R},n}}$ are defined in
Ref.~\onlinecite{pickard:2001} and satisfy
$\braket{\tilde{p}_{\vec{R},n}}{\tilde{\phi}_{\vec{R}', m}} =
\delta_{\vec{R},\vec{R}'}\delta_{n,m}$, where $\left\{\ket{\tilde{\phi}_{\vec{R},n}}\right\}$
is a set of pseudo partial-wavefunctions corresponding to the all-electron
partial wavefunctions $\left\{\ket{\phi_{\vec{R},n}}\right\}$.

The electronic magnetization operator $\bar{\vec{M}}(\vec{r'})$ is
reconstructed using PAW\cite{blochl:1994},
\begin{multline}
  \bar{\vec{M}}(\vec{r'}) = \sum_\sigma\rm{sgn}(\sigma)\frac{\vec{B}}{2cB}
      \Biggl\{\ket{\vec{r'}}\bra{\vec{r}'} \\ + 
      \sum_{n,m}\ket{\tilde{p_{\vec{R},n}}}\bigl[
      \braket{\Phi_{\vec{R},n}}{\vec{r}'}\braket{\vec{r'}}{\Phi_{\vec{R},m}} \\ -
      \braket{\tilde{\Phi}_{\vec{R},n}}{\vec{r}'}\braket{\vec{r}'}{\tilde{\Phi}_{\vec{R},m}} \bigr]
      \bra{\tilde{p}_{\vec{R},m}}\Biggr\}.\label{eq:psmag}
\end{multline}
To linear order in $B$, the spin and orbital response are
not coupled. Hence $\bar{\vec{M}}(\vec{r}')$ can be reconstructed using PAW
only, rather than the gauge including method GIPAW.

\section{Metallic System} 
\label{sec:smearing}
In order to treat the Fermi surface accurately and efficiently, we follow
Ref.~\onlinecite{gironcoli:1995} and introduce a fictitious temperature $T$ into the
electronic system. 

Let $\ket{\bar{\Psi}_i\zer}$ and $\epsilon_i\zer$ be the eigenvectors
and eigenvalues of the Hamiltonian $\bar{\mathcal{H}}\zer$ defined in
Eq.~\ref{eq:hamzer}.  Let $f(x)$ be a smooth step-function. The occupation
$f_{F,i}$ of energy level $i$ is defined as
$f_{F,i}=f(-\frac{\epsilon_F\zer-\epsilon_{i}\zer}{T})$, where
$\epsilon_F\zer$ is the Fermi energy. The latter is recovered from the
conservation of the number $2N$ of electrons in the system, $\sum_i
f_{F,i} = 2N$, with $i$ running over all eigenstates.

It was shown by de Gironcoli\cite{gironcoli:1995} that the first order
expectation value $o\one$ of an operator $\mathcal{O}  = \mathcal{O}\zer +
\mathcal{O}\one$, with $\zer$ ($\one$) indicating the zero (first) order
pertubation expansion, can be recovered as 
\begin{multline}
  o\one =  2\sum_{i\sigma}  \Re\left\{\bra{\bar{\Psi}_i\zer}\mathcal{O}\zer\ 
           \mathcal{G}(\epsilon_i\zer)\ 
           \bar{\mathcal{H}}\one \ket{\Psi_i\zer}\right\}\\
           + 2\sum_i f_{F,i}\bra{\bar{\Psi}_i\zer}\mathcal{O}\one\ket{\bar{\Psi}_i\zer}\\
           + 2\sum_i \frac{\epsilon_F\one}{T}\delta(-\frac{\epsilon_F\zer-\epsilon_i\zer}{T})
           \bra{\bar{\Psi}_i\zer}\mathcal{O}\zer\ket{\bar{\Psi}_i\zer}.\label{eq:expect}
\end{multline}
$\Re$ is the real value. The sum over $i$ runs over all states.
$\bar{\mathcal{H}}\one$ is some pertubation (it will be either the orbital or
spin pertubation of Eqs.~\ref{eq:hamorb} and \ref{eq:hamspin}). The last term
accounts for variations of the Fermi energy to first order $\epsilon_F\one$.
The linear variation of the Fermi energy can be recovered from the conservation of
the number of electrons,
$\sum_iT^{-1}(\epsilon_F\one-\epsilon_i\one)\delta(-\frac{\epsilon_F\zer-\epsilon_i\zer}{T})
= 0$. In this work we will always have $\epsilon_F\one=0$. Function
$\delta(x)$ is defined as the derivative of $f(x)$,
$\delta(x)=-\rm{d}f(x)/\rm{d}x$. The Green functions
$\mathcal{G}(\epsilon)$ is defined as 
\begin{equation}
  \label{eq:green}
  \bar{\mathcal{G}}(\epsilon_i) = \sum_j \frac{f_{F,j}-f_{F,i}}{\epsilon_j\zer - \epsilon_i\zer}
                            \ket{\bar{\Psi}_j\zer}\bra{\bar{\Psi}_j\zer}
\end{equation}
The sum over $j$ runs over all states. For $i=j$, the limit
$(f_{F,j}-f_{F,i})/(\epsilon_j\zer - \epsilon_i\zer)\overset{i=j}{\mapsto }
\delta(-\frac{\epsilon_F\zer-\epsilon_i\zer}{T})$ is taken. Expression
\ref{eq:expect} contains a factor two for spin.

\section{NMR orbital shifts}
\label{sec:orb}

The method presented in this section is an extension to metals of the scheme
proposed in Ref.~\onlinecite{pickard:2001} to compute NMR shifts in
insulators.  The Fermi surface is modeled using the smearing scheme of
Ref.~\onlinecite{gironcoli:1995}. For the sake of simplicity, the proof is
given for an all-electron system (i.~e.\ with $\mathcal{V}^{nl}_\vec{R}=0$).

We first compute the induced-current to first order for a finite system. The
result is re-expressed in a form suitable for extended systems using the
sum-rule of appendix \ref{appendix}. This expression is then specialized to the case
of periodic systems. Finally, we give the expression of the orbital current for
a pseudo-system.

\subsection{Finite systems}
By setting $\mathcal{V}^{nl}_\vec{R}=0$, the Hamiltonian of an all-electron
system is recovered from Eq.~\ref{eq:ham},
\begin{gather}
  \mathcal{H} = \mathcal{H}\zer + \mathcal{H}\one + O(B^2), \\
  \mathcal{H}\zer = \frac{1}{2}\vec{p}^2 + \mathcal{V}_{scf}(\vec{r}) \\
  \mathcal{H}_o\one = \frac{1}{2c}\vec{L} \cdot\vec{B}
\end{gather}
We note $\ket{\Psi_i}$ the all-electron wavefunctions.
The current operator for an all-electron system is given in
Eq.~\ref{eq:joperator}. Using the linear response Eq.~\ref{eq:expect}, the
expectation value $\vec{j}\one(\vec{r}')$ can be recovered as,
\begin{multline}
  \vec{j}\one(\vec{r}') = 2\sum_i \Re\left\{\bra{\Psi_i\zer}\vec{J}^p(\vec{r}')
        \ \mathcal{G}(\epsilon_i\zer)\ \mathcal{H}_o\one\ket{\Psi_i\zer}\right\} \\ 
        - \frac{\vec{B}\wedge\vec{r}'}{c}\rho\zer(\vec{r}').\label{eq:finite}
\end{multline}
In the above equation, we have used the assumption that there is no linear
order variation of the Fermi energy, $\epsilon_F\one=0$. Indeed, in a
non-degenerate system, the linear order variation of the eigenvalues are
$\epsilon_i\one = \bra{\Psi_i\zer}
\frac{1}{2c}\vec{L}\wedge\vec{B} \ket{\Psi_i\zer}$ for a given field $\vec{B}$.
Since the zero order system is invariant upon time reversal, the wave-functions
$\ket{\Psi_i\zer}$ can be chosen real. Hence, we have $\epsilon_i\one =0$. It
follows from the condition on $\epsilon_F\one$
given in section \ref{sec:smearing} that $\epsilon_F\one=0$.

Eq.~\ref{eq:finite} is valid for a finite system only. indeed, for
$r'\mapsto\infty$, $\vec{B}\wedge\vec{r}'\rho(\vec{r}')$ diverges in an extended
system. There is a similar divergence in the other term of Eq.~\ref{eq:finite},
such that the orbital current itself is finite. Yet, from a numerical point of view,
Eq.~\ref{eq:finite} cannot be used to compute $\vec{j}\one(\vec{r}')$.

\subsection{Extended System}
Following Ref.~\onlinecite{pickard:2001}, Eq.~\ref{eq:finite} can be
reexpressed using a generalized $f$-sum rule (given in appendix
\ref{appendix}) into a  more practical expression for an extended
system. We have, 
\begin{multline}
   \frac{\vec{B}\wedge\vec{r}'}{c}\rho\zer(\vec{r}') = \\
   \sum_if_{F,i}\bra{\Psi_i\zer} \frac{1}{c\imath}
          \left[\vec{B}\wedge\vec{r}'\cdot\vec{r}, \vec{J}^p(\vec{r})\right]\ket{\Psi_i\zer},
\end{multline}
where $\vec{J}^p$ is an odd operator and $\vec{B}\wedge\vec{r}'\cdot\vec{r}$ an
even operator. Using the sum rule, Eq.~\ref{eq:finite} can be rewritten as,
\begin{multline}
  \label{eq:extended}
  \vec{j}\one(\vec{r}') = \frac{1}{c}\sum_i \Re\biggl\{\bra{\Psi_i\zer}\vec{J}^p(\vec{r}') \\
         \mathcal{G}(\epsilon_i\zer)\ (\vec{r}-\vec{r}')\wedge\vec{p} \ket{\Psi_i\zer}\biggr\} 
\end{multline}
Since position quantities now enter as differences, it follows that the above
expressions is invariant upon translation of the system.  Furthermore, the
Green function at finite temperature is short-ranged. It follows that
contributions to the orbital current vanish for large values of
$(\vec{r}-\vec{r}')$ in Eq.~\ref{eq:extended}. 

\subsection{Periodic System}
At this point, we have a formalism adequate for obtaining the current response
in extended metallic systems. Of those, only translationally-invariant periodic systems
are computationally feasible. Hence, we now introduce these translational
symmetries explicitly into the equations for
the current response. We write $\ket{\Psi_{i\vec{k}}\zer} =
e^{\imath\vec{k}\cdot\vec{r}}\ket{u_{i\vec{k}}\zer}$ the electronic Bloch
states of crystal momentum $\vec{k}$. $\epsilon_{i\vec{k}}\zer$ is the
corresponding eigenvalue. $\braket{\vec{r}}{u_{i\vec{k}}\zer}$ is a normalized
cell-periodic function. In the spirit of Ref.~\onlinecite{pickard:2001}, we
transform the real-space dependence $(\vec{r}-\vec{r}')$ into a reciprocal
space dependence by introducing the limit,
\begin{equation}
  (\vec{r}-\vec{r}') = \lim_{q\mapsto 0}\frac{1}{2q}\sum_{\alpha=x,y,z}
                      \left[e^{\imath q\vec{u}_\alpha\cdot(\vec{r}-\vec{r}')} -
                            e^{-\imath q\vec{u}_\alpha\cdot(\vec{r}-\vec{r}')} \right],
\end{equation}
where $\vec{u}_{\alpha=x,y,z}$ is real-space basis. This transformation is
subject to the condition $|\vec{r}-\vec{r}'|<\vec{C}$ ($\vec{C}$ a vector) which
is verified since contributions to the orbital current in Eq.~\ref{eq:extended} vanish for
large values of $(\vec{r}-\vec{r}')$.  The orbital current is then recovered as
a numerical derivative,
\begin{gather}
  \vec{j}\one(\vec{r}') = \lim_{q\mapsto0}\frac{1}{2q}
            \left[\vec{S}(\vec{r}',q) - \vec{S}(\vec{r}',-q)\right], \label{eq:periodic}
\end{gather}
where,
\begin{multline}
  \vec{S}(\vec{r}',q) = \frac{1}{cN_k}\sum_{\alpha=x,y,z}\sum_{i,\vec{k}}
         \Re\Biggl\{\frac{1}{\imath}
         \bra{u_{i\vec{k}}\zer}\vec{J}^p_{\vec{k},\vec{k}+q\vec{u}_\alpha}(\vec{r}')\\
          \ \mathcal{G}_{\vec{k}+q\vec{u}_\alpha}(\epsilon_{i\vec{k}})\  
          \vec{B}\wedge\vec{u}_\alpha\cdot(\vec{p}+\vec{k})
          \ket{u_{i\vec{k}}\zer}\Biggr\}.
\end{multline}   
$N_k$ is the number of $\vec{k}$-points in the discrete integration of the Brillouin zone.
We have introduced the $\vec{k}$-dependent Green function $\mathcal{G}_\vec{k}(\epsilon)$,
\begin{equation}
  \mathcal{G}_\vec{k}(\epsilon)
         = \sum_j \frac{f_{F,j\vec{k}}- f\left(-\frac{\epsilon_F\zer-\epsilon}{T}\right)}
                       {\epsilon_{j\vec{k}}\zer-\epsilon}\ket{u_{j\vec{k}}\zer}
                       \bra{u_{j\vec{k}}\zer},
\end{equation}
and the $\vec{k}$-dependent paramagnetic current operator
$\vec{J}^p_{\vec{k},\vec{k'}}$,
\begin{equation}
  \vec{J}^p_{\vec{k},\vec{k}'}(\vec{r}') = 
        -\frac{1}{2}\left(\vec{p}+\vec{k}\right)\ket{\vec{r}'}\bra{\vec{r}'}
        -\frac{1}{2}\ket{\vec{r}'}\bra{\vec{r}'}\left(\vec{p}+\vec{k}'\right)
\end{equation}

Eq.~\ref{eq:periodic} allows us to compute the orbital current of an
all-electron system. In practice, it is more efficient to use pseudo-potentials
when expanding the density on a plane-wave basis set. We now give a general
expression for the orbital current in periodic pseudo-potential systems using the
GIPAW reconstruction scheme of Ref.~\onlinecite{pickard:2001}.

\subsection{Periodic pseudo-potential system}

The orbital current can be obtained from a pseudo-system using Eq.~\ref{eq:ham}
and Eq.~\ref{eq:curps}. Following Ref.~\onlinecite{pickard:2001} as well as the
steps given above, one can find an expression for the orbital current suited to
a periodic pseudo-system.

We find that the current is composed of three components: (i) the bare
current $\vec{j}\one_{bare}(\vec{r}')$, (ii) the paramagnetic augmentation
current $\vec{j}\one_{\Delta p}(\vec{r}')$, (iii) the diamagnetic augmentation
current $\vec{j}_{\Delta d}\one(\vec{r}')$.
\begin{equation}
  \vec{j}\one(\vec{r}') =   \vec{j}\one_{bare}(\vec{r}') 
                          + \vec{j}\one_{\Delta p}(\vec{r}')
                          + \vec{j}\one_{\Delta d}(\vec{r}')
\end{equation}

The diamagnetic augmentation current is simply the expectation value of the
operator given in Eq.~\ref{eq:diam},
\begin{equation}
  \vec{j}_{\Delta d}\one(\vec{r}') = 2\sum_{i,\vec{R}}
       \bra{\bar{\Psi}_{i\vec{k}}\zer} \Delta\vec{J}^d_\vec{R}(\vec{r}')
       \ket{\bar{\Psi}_{i\vec{k}}\zer}
\end{equation}
Note that the projectors $\ket{\tilde{p}_{n,\vec{R}}}$ make
$\Delta\vec{J}^d_\vec{R}(\vec{r}')$ short-ranged. Furthermore, since positions
quantities enter as differences, $\vec{j}_{\Delta d}\one(\vec{r}')$ is
translationally invariant.

The paramagnetic augmentation and bare currents are obtained as numerical
differences,
\begin{gather}
  \vec{j}\one_{bare}(\vec{r}') = \lim_{q\mapsto0}\frac{1}{2q}
                                 \left[\vec{S}_{bare}(\vec{r}',q) 
                                       - \vec{S}_{bare}(\vec{r}',-q)\right], \label{eq:bareper}\\
  \vec{j}\one_{\Delta p}(\vec{r}') = \lim_{q\mapsto0}\frac{1}{2q}
                                 \left[\vec{S}_{\Delta p}(\vec{r}',q) 
                                       - \vec{S}_{\Delta p}(\vec{r}',-q)\right]. \label{eq:paraper}
\end{gather}
The two newly introduced functions are defined as,
{\small
\begin{multline}
  \vec{S}_{bare}(\vec{r}',q) = \frac{1}{cN_k}\sum_{\alpha=x,y,z}\sum_{i,\vec{k}}
         \Re\Biggl\{\frac{1}{\imath}
         \bra{\bar{u}_{i\vec{k}}\zer}\vec{J}^p_{\vec{k},\vec{k}+q\vec{u}_\alpha}(\vec{r}')\\
         \ \bar{\mathcal{G}}_{\vec{k}+q\vec{u}_\alpha}(\epsilon_{i\vec{k}})\  
                \vec{B}\wedge\vec{u}_\alpha\cdot
                \vec{v}_{\vec{k}+q\vec{u}_\alpha,\vec{k}}\ket{\bar{u}_{i\vec{k}}\zer}\Biggr\},
\end{multline}}
and,
{\small
\begin{multline}
  \vec{S}_{\Delta p}(\vec{r}',q) = \frac{1}{cN_k}\sum_{\alpha=x,y,z}\sum_{i,\vec{k}}
         \Re\Biggl\{\frac{1}{\imath}\bra{\bar{u}_{i\vec{k}}\zer}
         \Delta\vec{J}^p_{\vec{L},\vec{\tau},\vec{k},\vec{k}+q\vec{u}_\alpha}(\vec{r}')\\
         \ \bar{\mathcal{G}}_{\vec{k}+q\vec{u}_\alpha}(\epsilon_{i\vec{k}})\  
              \vec{B}\wedge\vec{u}_\alpha\cdot
              \vec{v}_{\vec{k}+q\vec{u}_\alpha,\vec{k}}\ket{\bar{u}_{i\vec{k}}\zer}\Biggr\}.
\end{multline}}
$\ket{\bar{u}_{i\vec{k}}\zer}$ is the cell-periodic function such that
$\ket{\bar{\Psi}_{i\vec{k}}\zer}=e^{\imath\vec{k}\wedge\vec{r}}\ket{\bar{u}_{i\vec{k}}\zer}$.
The Green function  $\bar{\mathcal{G}}_\vec{k}(\epsilon)$ is redefined using
the pseudo-eigenstates,
\begin{equation}
  \bar{\mathcal{G}}_\vec{k}(\epsilon_{i\vec{k}}\zer)
           = \sum_j \frac{f_{F,j\vec{k}}- f\left(-\frac{\epsilon_F\zer-\epsilon}{T}\right)}
                         {\epsilon_{j\vec{k}}\zer-\epsilon}\ket{\bar{u}_{j\vec{k}}\zer}
                         \bra{\bar{u}_{j\vec{k}}\zer},
\end{equation}

A $\vec{k}$-dependent non-local pseudopotential
$\mathcal{V}^{nl}_{\vec{k},\vec{k}'}$ is also defined, which acts on $\vec{k}$
Bloch states on the left and $\vec{k}'$ states on the right,
\begin{equation}
  \mathcal{V}^{nl}_{\vec{k},\vec{k}'} = \sum_{\vec{\tau}}\sum_{n,.m}
          \ket{\tilde{p}_{\vec{\tau},n}^\vec{k}}
          a^\tau_{n,m}\bra{\tilde{p}_{\vec{\tau},n}^{\vec{k}'}}.
\end{equation}
The periodic projectors $\ket{\tilde{p}_{\vec{\tau},n}^\vec{k}}$ are obtained
from the real-space projectors $\ket{\tilde{p}_{\vec{L}+\vec{\tau},n}}$ as
\begin{equation}
  \ket{\tilde{p}_{\vec{\tau},n}^\vec{k}} = \sum_\vec{L}
           e^{-\imath\vec{k}\cdot(\vec{r}-\vec{L}-\vec{\tau})}
           \ket{\tilde{p}_{\vec{L}+\vec{\tau},n}},
\end{equation}
where the sum runs over the lattice vectors $\vec{L}$. 
Cell-internal atomic-coordinates are noted with $\vec{\tau}$. The velocity operator is also redefined as,
\begin{equation}
  \vec{v}_{\vec{k},\vec{k}'} = \vec{p} + \vec{k}'
        + \frac{1}{\imath}\left[\vec{r}, \mathcal{V}^{nl}_{\vec{k},\vec{k}'}\right].
\end{equation}
Finally, a $\vec{k}$-dependent paramagnetic current operator
$\vec{J}^p_{\vec{k},\vec{k}'}$ and its affiliate pseudo-operator
$\Delta\vec{J}^p_{\vec{L},\vec{\tau},\vec{k},\vec{k}'}$ are introduced.
\begin{equation}
  \vec{J}^p_{\vec{k},\vec{k}'}(\vec{r}') = 
        -\frac{1}{2}\left(\vec{p}+\vec{k}\right)\ket{\vec{r}'}\bra{\vec{r}'}
        -\frac{1}{2}\ket{\vec{r}'}\bra{\vec{r}'}\left(\vec{p}+\vec{k}\right)
\end{equation}
\begin{multline}
  \Delta\vec{J}^p_{\vec{L},\vec{\tau},\vec{k},\vec{k}'} 
         = \sum_{n,m} \ket{\tilde{p}_{\vec{\tau},n}^\vec{k}}\bigl[
                      \bra{\phi_{\vec{L}+\vec{\tau},n}}\vec{J}^p(\vec{r}')
                      \ket{\phi_{\vec{L}+\vec{\tau},m}}\\
                      - \bra{\tilde{\phi}_{\vec{L}+\vec{\tau},n}}\vec{J}^p(\vec{r}')
                        \ket{\tilde{\phi}_{\vec{L}+\vec{\tau},m}}
                      \bigr]\bra{\tilde{p}_{\vec{\tau},n}^\vec{k}}
\end{multline}

The orbital shielding is then obtained from Eq.~\ref{eq:field} and from its
definition $\vec{B}_o = -\tensor{\sigma}_o\cdot\vec{B}$.

\section{Knight Shift}
\label{sec:knight}
We now turn to the Knight shift, which results from the electrons
interacting with the field as spinning charges. More specifically, the magnetic
field induces a net electronic-spin which then interacts with the magnetic nuclear
dipole through the Hyperfine interaction (Eq.~\ref{eq:dipolefield}). The Knight
shift measures this interaction.

The Hamiltonian to first order is given up to first order be Eq.~\ref{eq:ham},
Eq.~\ref{eq:hamzer}, and Eq.~\ref{eq:hamspin},
\begin{gather}
  \mathcal{H}\zer = \frac{1}{2}\vec{p}^2
                          + \mathcal{V}\zer_{scf} \\
  \mathcal{H}_{s\sigma}\one = \rm{sgn}(\sigma)B\left(\frac{g_e}{4c}
  +\mathcal{V}\one_{scf}\right),
\end{gather}
The linear order wavefunctions $\ket{\Psi_{i\sigma}\one}$ and eigenvalues
$\epsilon_{i\sigma}\one$ are anti-symmetric with respect to field direction,
i.~e.\ when $\vec{B}$ is mapped onto $\vec{B}\mapsto-\vec{B}$, we expect
$\epsilon_{i\sigma}\one\mapsto\epsilon_{i\bar{\sigma}}\one$ and
$\ket{\Psi_{i\sigma}\one}\mapsto\ket{\Psi_{i\bar{\sigma}}\one}$,
where $\bar{\sigma}$ is the spin opposite to $\sigma$. It follows then that
$\ket{\Psi_{i\uparrow}\one}=-\ket{\Psi_{i\downarrow}\one}$ and
$\epsilon_{i\uparrow}\one = -\epsilon_{i\downarrow}\one$. From this last
condition, it follows that there is no variation of the Fermi energy to first
order, $\epsilon_F\one=0$.

For simplicity, the following is obtained directly for the
pseudo-system. Indeed, the reconstruction of the constant part of
$\mathcal{H}\one_\sigma$ is zero. Furthermore, we neglect the polarization of
the core electrons by the valence spin-density. In practice, this is equivalent
to neglecting the PAW reconstruction of the self-consistent pertubation.

 Exploiting the spin anti-symmetry described above, the electronic
 magnetization to fist order in $B$ can be obtained as, 
\begin{multline}
  \bar{\vec{m}}\one(\vec{r'}) = 2\sum_i
  \Re\Biggl\{\bra{\bar{\Psi}_i\zer}\vec{\bar{M}}(\vec{r}')\\
  \bar{\mathcal{G}}(\epsilon_i\zer)\ \frac{1}{2} \left(
        \mathcal{H}_{s\uparrow}\one-\mathcal{H}_{s\downarrow}\one \right)
        \ket{\bar{\Psi}_i\zer}\biggr\}\label{eq:mps},
\end{multline}
with the quantities defined previously. 

Once the electronic magnetization is obtained, the Knight shift can be
computed from Eq.~\ref{eq:dipolefield} and
Eq.~\ref{eq:b}.

\section{Practical implementation}
\label{sec:practical}
The goal of the method presented above is to provide a practical and quantitative
approach to computing NMR shifts in metals. It was implemented in a
parallel plane-wave pseudopotential electronic structure code. We now
outline the features specific to the NMR method. We shall first discuss the
application of the Green function, common to both orbital and Knight shift
computations, and then turn to the specifics of each type of response.

\subsection{Linear response} 
we are interested in computing first-order quantities (see
Eqs.~\ref{eq:bareper}, \ref{eq:paraper}, and \ref{eq:mps}) such as,
\begin{equation}
  o\one = \sum_i \bra{\bar{\Psi}_i\zer}\mathcal{O}\ \bar{\mathcal{G}}(\epsilon_i\zer)\ 
  \mathcal{H}\one \ket{\bar{\Psi}\one_i},
\end{equation}
where $\mathcal{O}$ is an operator and $\mathcal{H}\one$ some pertubation. the
green function is expressed as in Eq.~\ref{eq:green}. Both the sum over $i$
above, and that over $j$ in $\bar{\mathcal{G}}(\epsilon)$ range over all states. such
an expression cannot be calculated directly. it was shown by de Gironcoli in
ref.~\onlinecite{gironcoli:1995} that $o\one$ can be computed \textit{via} an
alternate first-order wavefunction $\ket{\delta\bar{\Psi}_i\one}$,
\begin{equation}
  o\one = 2\sum_i
  \Re\left\{\bra{\bar{\Psi}_i\zer}\mathcal{O}\ket{\delta\bar{\Psi}_i\one}\right\},
\end{equation}
such that the sum over $i$ runs only over partially occupied states.
$\ket{\delta\Psi_i\one}$ can be also computed without reference to empty
states. 
\begin{gather}
      \left[\mathcal{H}\zer + \mathcal{Q}
      -\epsilon_i\zer\right]\ket{\delta\bar{\Psi}_i\one} =
      -\left[f_{F,i}-\aleph_i\right]\mathcal{H}\one\ket{\bar{\Psi}_i\zer}\notag \\
      \mathcal{Q} = \sum_j \alpha_j \ket{\bar{\Psi}_j\zer}\bra{\bar{\Psi}_j\zer},
      \quad \aleph_i =
      \sum_j\beta_{i,j}\ket{\bar{\Psi}_j\zer}\bra{\bar{\Psi}_j\zer} \notag\\
      \alpha_j = \max\left(\epsilon_F\zer + nT - \epsilon_j\zer,
      0\right) \notag\\
      \beta_{i,j} = f_{F,i}g_{i,j} + f_{F,j}g_{j,i} 
                    + \alpha_j\frac{f_{F,i}-f_{F,j}}{\epsilon_i\zer-\epsilon_j\zer}g_{j,i}
   \label{eq:system}
\end{gather}
$g(x)$ is a symmetric function such that $g(x)+g(-x)=1$. We define $g_{ij} =
g(\frac{\epsilon_i\zer-\epsilon_j\zer}{\sigma})$. Partially occupied
wavefunctions are defined such that $\epsilon_i\zer < \epsilon_F\zer - nT<0$
($\alpha_i \neq 0$), where n is a suitably large number. We find that orbital
and spin shieldings are converged for $n=7$.

\subsection{Orbital shifts}
The method presented above differs only slightly from the prior method for
insulators. We will address only these differences and defer the interested
reader to Ref.~\onlinecite{pickard:2001}. 

The macroscopic induced field $\vec{B}_o\one(\vec{G}=0)$, where $\vec{G}$ is a
vector of reciprocal space, is not a bulk property. Indeed it results from the
surface current in the sample, and hence depends on the shape of the sample.
Following Ref.~\onlinecite{pickard:2001}, we compute it through the so-called
\emph{bare} macroscopic susceptibility $\tensor{\chi}_{bare}$, consistent with the on-site approximations for the reconstruction current,
\begin{equation}
  \vec{B}\one_o(\vec{G}=0) = \frac{2}{3}4\pi \tensor{\chi}_{bare}\cdot\vec{B},
\end{equation}
$\tensor{\chi}_{bare}$ is the contribution to the macroscopic susceptibility from
the bare current $\vec{j}\one_{bare}$. We adapt the ansatz of
Ref.~\onlinecite{pickard:2001} to the case of metallic compounds,
\begin{equation}
  \tensor{\chi}_{bare} = \lim_{q\mapsto0}\frac{1}{q^2}\left[\tensor{F}(q)+\tensor{F}(-q) -2 \tensor{F}(0)\right],
\end{equation}
where $F_{ij}=(2-\delta_{ij})Q_{ij}(q)$. $i$ and $j$ are Cartesian indices. 
{\small
\begin{multline}
  \tensor{Q}(q) = - \frac{1}{2c^2N_k\Omega}\sum_{\alpha=x,y,z}\sum_{i,\vec{k}}\Re\biggl\{\\
                  \bra{\bar{u}_{i\vec{k}}\zer} \vec{u}_\alpha\wedge\left(\vec{p}+\vec{k}\right)\ 
                  \bar{\mathcal{G}}_{\vec{k}+q\vec{u}_\alpha}(\epsilon_{i\vec{k}}\zer)\
                  \vec{u}_\alpha\wedge\vec{v}_{\vec{k}+q\vec{u}_\alpha,\vec{k}}
                  \ket{\bar{u}_{i\vec{k}}\zer}\biggr\}
\end{multline}}

When interested specifically in the susceptibility $\tensor{\chi_o}$, we use
another ansatz from Ref.~\onlinecite{pickard:2001}, with
\begin{equation}
  \tensor{\chi}_o = \lim_{q\mapsto0}\frac{1}{q^2}
  \left[\tensor{F}^{tot}(q)+\tensor{F}^{tot}(-q) -2 \tensor{F}^{tot}(0)\right],
\end{equation}
{\small
\begin{multline}
  \tensor{Q}^{tot}(q) = - \frac{1}{2c^2N_k\Omega}\sum_{\alpha=x,y,z}\sum_{i,\vec{k}}\Re\biggl\{\\
                  \bra{\bar{u}_{i\vec{k}}\zer} \vec{u}_\alpha\wedge
                       \vec{v}_{\vec{k},\vec{k}+q\vec{u}_\alpha}\\
                  \bar{\mathcal{G}}_{\vec{k}+q\vec{u}_\alpha}(\epsilon_{i\vec{k}}\zer)\
                  \vec{u}_\alpha\wedge\vec{v}_{\vec{k}+q\vec{u}_\alpha,\vec{k}}
                  \ket{\bar{u}_{i\vec{k}}\zer}\biggr\}
\end{multline}}
and $F^{tot}_{ij}=(2-\delta_{ij})Q^{tot}_{ij}(q)$\\

At zero temperature, $\tensor{\chi}_o$ and $\tensor{\chi}_{bare}$ above and the
corresponding quantities of Ref.~\onlinecite{pickard:2001} are equivalent.\\

\subsection{Knight shift}
The variation of the self-consistent potential
$\mathcal{V}_{scf}\one(\vec{r})$ is evaluated using a simple self-consistent
loop over the calculation of the first-order wave-functions.  In other words,
the spin density is recomputed at each step and
$\mathcal{V}_{scf}\one(\vec{r})$ updated. In the case of local density
approximations, $\mathcal{V}_{scf}\one(\vec{r})$ is simply,
\begin{equation}
  \begin{split}
    \mathcal{V}\one_{scf}(\vec{r}) &= 
           \frac{\partial \mathcal{V}_{scf}(\vec{r})}
                {\partial\rho_s(\vec{r})}\rho_s\one(\vec{r}), \\
   &= \frac{\partial }{\partial\rho_s(\vec{r})}
      \left(\mathcal{V}_{xc}^\uparrow(\vec{r})-\mathcal{V}_{xc}^\downarrow(\vec{r})\right)
      \rho_s\one(\vec{r}), \\
  \end{split}
\end{equation}
where $V_{xc}^\uparrow$ and $V_{xc}^\downarrow$ are the
exchange-correlation potential of the up and down spin channels,
respectively, computed from the ground state densities.
$\rho_s\one(\vec{r})=\rho_\uparrow\one(\vec{r})-\rho_\downarrow\one(\vec{r})$ is
the first order spin density at $\vec{r}$. These derivatives are
evaluated numerically for each point of the real space mesh. The
self-consistent Hartree potential is not spin dependent, and hence it is not
modified by variations of the spin density.

A pertubation using generalized gradient approximations can be implemented in
much the same way.

We find that  convergence with respect to the number of iterations over
$\mathcal{V}_{scf}\one(\vec{r})$ can be achieved efficiently without mixing.

\section{Numerical Tests}
\label{sec:tests}

\subsection{Interacting Homogeneous gas}

NMR shifts require the computation of the macroscopic susceptibility in order
to account for the diamagnetic shielding resulting from surface currents. We
will now test these calculations against available analytical results for the
homogeneous electron gas.  The orbital ($\chi_o$) and spin ($\chi_s$) 
susceptibilities per unit volume of this model system are given by the
formul\ae\cite{ashcroft}:
\begin{equation}
  \begin{aligned}
    \chi_o &= -\frac{1}{12c^2} g(\epsilon_F),\\
    \chi_s &= -\frac{g_e}{4c^2}g(\epsilon_F)\frac{1}{1+g(\epsilon_F)\frac{\partial^2\epsilon_{xc}}{\partial\rho_s^2}},\\
    g(\epsilon_F) &= \left(\frac{3}{\pi^4}\frac{N}{\Omega}\right)^{1/3}.\\
  \end{aligned}\label{eq:III.4-suscept}
\end{equation}
$N$ is the number of electrons in  the system, $\epsilon_{xc}$ is the
exchange-correlation energy per unit volume as given by PBE\cite{perdew:1996},
and $g(\epsilon_F)$ is the density of states at the Fermi energy. The
derivative of $\epsilon_{xc}$ is evaluated numerically.

The fractional factor in $\chi_s$ results from the exchange-correlation. More
specifically, the magnetic field induces a polarization of the electrons at the
Fermi energy, which then propagates to lower lying levels through
exchange-correlation interactions. Indeed, for a non-interacting homogeneous gas,
the spin susceptibility reduces to $\chi_s = g_e/(4c^2) g(\epsilon_F)$, i.~e.\
it is simply proportional to the available degrees of freedom at the Fermi
surface.  This propagation effect is rendered computationally by the
self-consistency of Eq.~\ref{eq:mps}. 

To simulate a homogeneous gas within a pseudo-potential code, we construct a
pseudopotential with zero potential and zero atomic charge. A temperature of
0.4\,eV is introduced into the system. We use an fcc unit cell with
a cell-parameter of 3.61\AA.\ The Brillouin zone is sampled with a
60x60x60 Monkhorst-Pack grid. Different electronic densities are
obtained by varying the number of electrons in the cell.

\begin{figure}[htbp]
  \begin{center}
    \vspace{0.5cm}
    \includegraphics[width=\columnwidth]{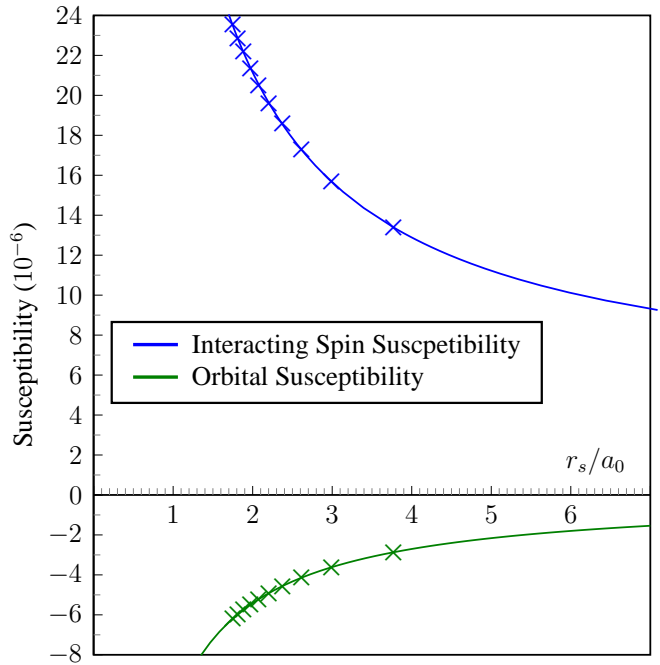}
    \caption{Spin and Orbital susceptibility per unit
      volume of the interacting-electron homogeneous gas respect to
      $r_s/a_0=(3/4\pi\rho)^3$.  Solid lines represent analytical
      results while dots represent results computed with this code at
      $0.4$\,eV smearing.  The susceptibilities are dimensionless. The
      exchange-correlation is modeled with the PBE functional. }
    \label{fig:gas}
  \end{center}
\end{figure}

Results are given for a range of densities (parameterized by $r_s/a_o =
(3/4\pi\rho)^3$, where $a_0$ is Bohr constant) in Fig.~\ref{fig:gas}. X dots
represent the response computed with our approach, and solid
lines are analytical results. The PBE\cite{perdew:1996} exchange-correlation
functional is used. Results agree to within numerical noise.

\section{Simple metals}

The object of the present work is to build a quantitative method for computing
NMR shifts in metallic compounds. As such, we now study three simple metals:
bulk aluminum, bulk lithium, and bulk copper.

Experimentally, NMR shifts are obtained with respect to the response of
so-called zero-shift compounds. We will first study this aspect of the problem,
and compute the shielding of these compounds. We will then give the
computational details for each metal, and finally examine the NMR shifts and 
macroscopic magnetic susceptibilities.

\subsection{Computational Details}

  Computational details are reported in tables \ref{tab:params}.  For all
  calculations, we use the Marzari-Vanderbilt smearing
  function\cite{marzari:1999} and Troullier-Martins\cite{troullier:1991}
  norm-conserving pseudopotentials.  Following experimental conventions, we use
  a spherical sample when accounting for surface currents. We use the
  PBE\cite{perdew:1996} exchange-correlation functional.

  Aluminum and copper are cubic face centered metals with $a=4.05$\,\AA{} and
$a=3.61$\,\AA{}, respectively. Lithium is body centered with $a=3.49$\,\AA{}. We
use experimental cell parameters as given by Ref.~\onlinecite{ashcroft}.

\begin{table}[tbph]
  \begin{center}
     \begin{ruledtabular}
     \begin{tabular}{c|ccccc}
        \multirow{2}*{Metal} & \multirow{2}*{Smearing} & \multicolumn{2}{c}{Cutoff} & \multicolumn{2}{c}{$N_k$}\\
        & & Knight & Orbital & Knight & Orbital \\\hline
        Al & 0.15 & 15 & 15 & 29820 & 62790 \\
        Li &  0.2 & 15 & 15 &  8094 & 11900 \\
        Cu &  0.2 & 75 & 90 &  1300 &  5740 \\
     \end{tabular}
     \end{ruledtabular}
     \caption{Computational Details. Smearings are given in eV.
     Plane wave cutoffs for both Knight and orbital shift calculations
     are given in Rydberg.  $N_k$ stands for the number of independent
     $\vec{k}$-points in the irreducible wedge of the Brillouin zone.
     The Brillouin zone is represented with a discrete Monkhorst-Pack
     grid\cite{monkhorst:1976}. The Marzari-Vanderbilt smearing
     function\cite{marzari:1999} is used.}
     \label{tab:params}
  \end{center}
\end{table}

\subsection{Zero shift compounds}

Experimental NMR shifts are obtained as
\begin{equation}
  \vec{u}\cdot\tensor{\delta}\cdot\vec{u} = -\vec{u}\cdot(\tensor{\sigma}-\tensor{\sigma}_{ref})\cdot\vec{u}\label{eq:sigref},
\end{equation}
where $\vec{u}$ is the direction in which the external magnetic field is
applied, $\tensor{\sigma}$ is the the shielding of the compound, and
$\tensor{\sigma}_{ref}$ is the shielding of the zero-shift compound. 

Rather than evaluating $\tensor{\sigma}_{ref}$ directly, we will compute the
shielding of some compound for which the NMR shift is well known
experimentally, and then deduce $\tensor{\sigma}_{ref}$ from
Eq.~\ref{eq:sigref}. 

\begin{table}[tbph]
  \begin{center}
     \begin{ruledtabular}
     \begin{tabular}{c|c|ccc|c}
        Atom & ``zero-shift compound'' & \multicolumn{3}{c|}{compound} & $\sigma^{ref}$ \\
             &  & type & $\sigma^{th}$ & $\delta_{exp}$ & \\ \hline
        Al & AlCl$_3$ in heavy water & AlPO$_4$ & 519 & 45 & 564 \\
        Li & aqueous LiCl & Li$_2$O & 86 & 10 & 96 \\
        Cu & CuBr powder & CuBr & 424 & 0.0 & 424 \\
     \end{tabular}
     \end{ruledtabular}
     \caption{Reference shifts $\sigma^{ref}$. The ``compound'' columns gives
              the solid which is used to obtain the reference shift, its calculated isotropic
              shielding $\sigma^{th}$, and its experimental isotropic shift $\delta_{exp}$. The
              reference shift is obtained using the relationship $\delta_{exp} =
              -(\sigma^{th}-\sigma^{ref})$. Shieldings are converged to better than a ppm.
              Shieldings and shifts are given in ppm.}
     \label{tab:refs}
  \end{center}
\end{table}

The reference shifts for each element Al, Li, and Cu are given in
Tab.~\ref{tab:refs}. Note that all references are computed on insulators, and
hence that the shieldings result only from the orbital response. The latter are
computed using the method for insulators described in Ref.~\onlinecite{pickard:2001}.

\subsection{Behavior with respect to smearing}

\begin{figure}[htbp]
  \begin{center}
    \vspace{0.5cm}
    \includegraphics[width=0.85\columnwidth]{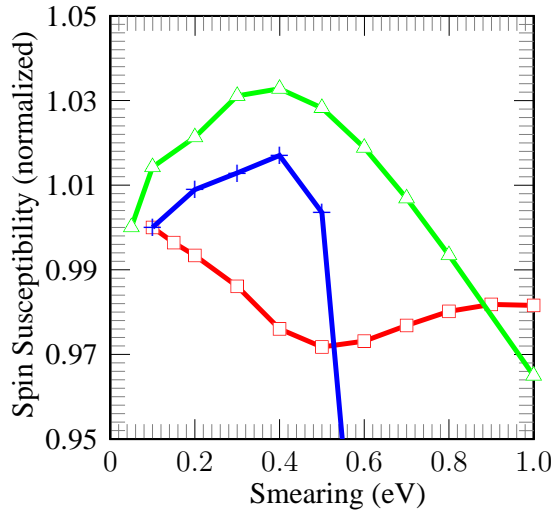}
    \caption{Convergence with respect to smearing ($\sigma$) of the spin
    susceptibility of Aluminum (green triangles), Lithium (red squares), and Copper (blue +).
    For comparison purposes, the spin susceptibility of each metal is
    normalized to its value at the lowest achieved smearing. The
    $x$-axis represents smearing in eV. }
    \label{fig:suscept-knight}
  \end{center}
\end{figure}

\begin{figure}[htbp]
  \begin{center}
    \vspace{0.5cm}
    \includegraphics[width=0.85\columnwidth]{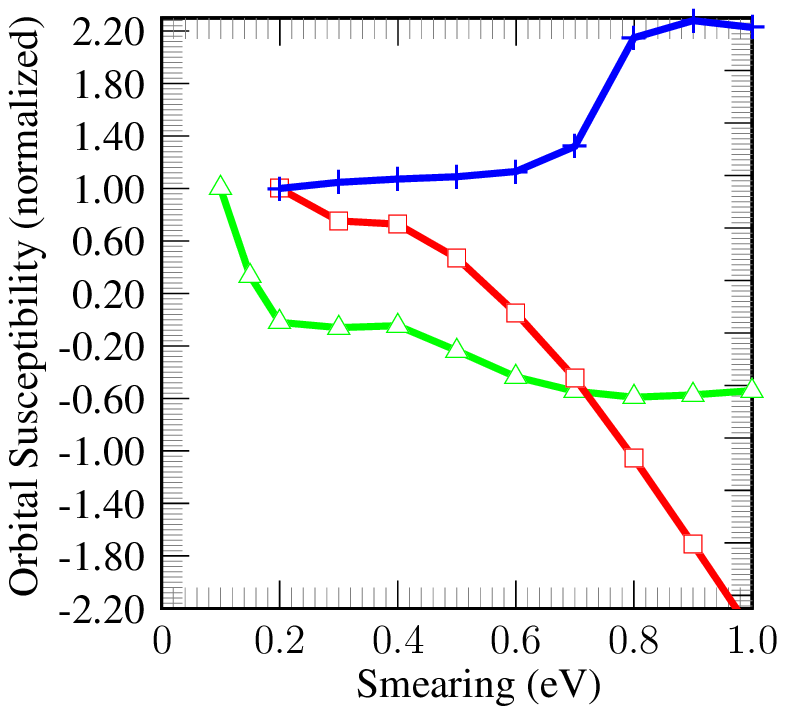}
    \caption{Convergence with respect to smearing ($\sigma$) of the
    orbital susceptibility of Aluminum (green triangles), Lithium (red
    squares), and Copper (blue +).  For comparison purposes, the
    orbital susceptibility of each metal is normalized to its value at the lowest
    achieved smearing. The $x$-axis represents smearing
    in eV. }
    \label{fig:suscept-orb}
  \end{center}
\end{figure}

\begin{figure}[htbp]
  \begin{center}
    \vspace{0.5cm}
    \includegraphics[width=0.85\columnwidth]{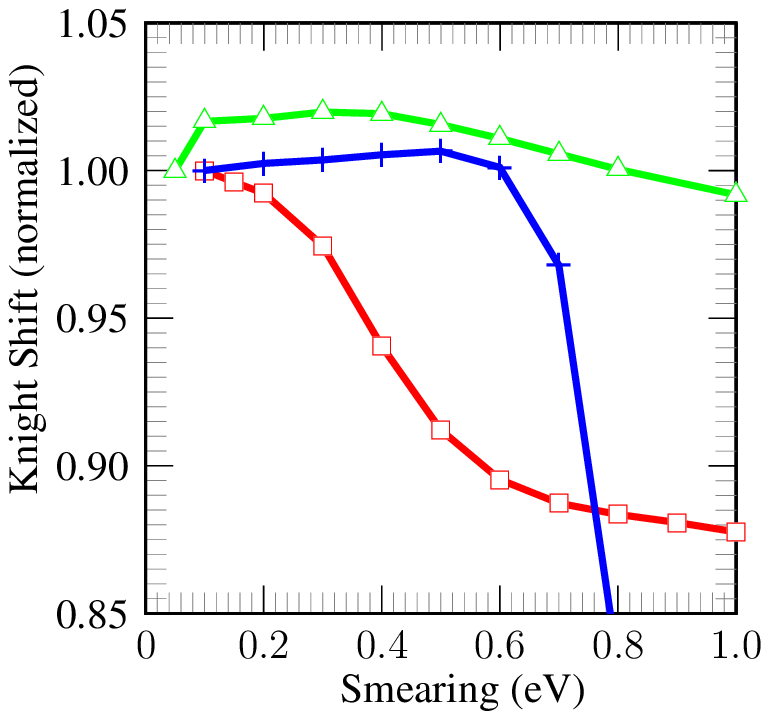}
    \caption{Convergence with respect to smearing ($\sigma$) of the Knight
    shift (not including the spin susceptibility) of Aluminum (green triangles),
    Lithium (red squares), and Copper (blue +).  For comparison
    purposes, the Knight shift of each metal is normalized to its value at the
    lowest achieved smearing. The $x$-axis
    represents smearing in eV. }
    \label{fig:shift-knight}
  \end{center}
\end{figure}

\begin{figure}[htbp]
  \begin{center}
    \vspace{0.5cm}
    \includegraphics[width=0.85\columnwidth]{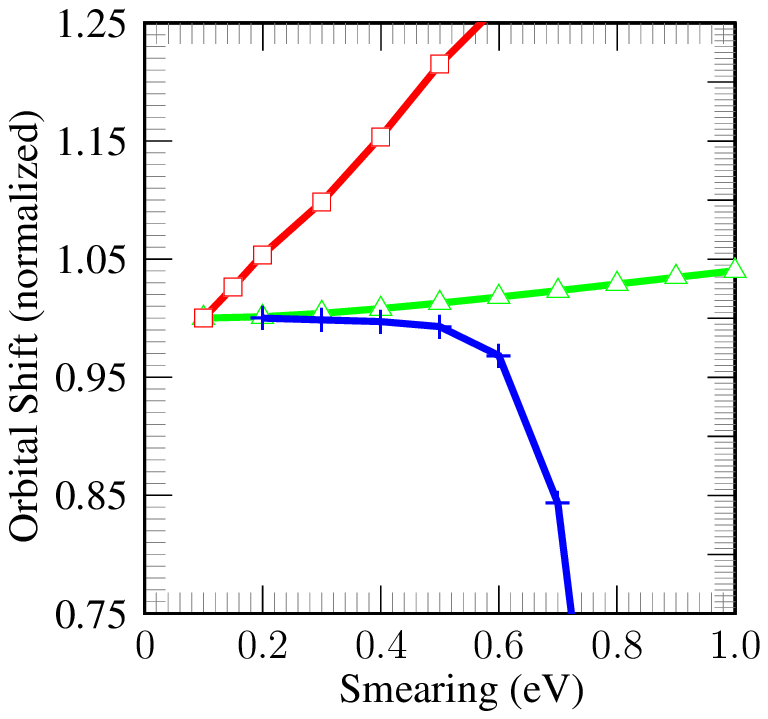}
    \caption{Convergence with respect to smearing ($\sigma$) of the orbital
    shift (not including the orbital susceptibility) of Aluminum (green triangles),
    Lithium (red squares), and Copper (blue +).  For comparison
    purposes, the orbital shift of each metal is normalized to its
    value at the lowest achieved smearing. The $x$-axis
    represents smearing in eV. }
    \label{fig:shift-orb}
  \end{center}
\end{figure}

The computation of NMR shifts requires a very fine description of the Fermi
surface. Hence, one must take care that the computed shifts are indeed
converged with respect to smearing. Figures \ref{fig:suscept-knight} and
\ref{fig:suscept-orb} report the convergence behavior with respect to smearing of, respectively, the spin
macroscopic spin-susceptibility, and of the macroscopic orbital-susceptibilities for
Aluminum, Lithium, and Copper. Figures \ref{fig:shift-knight} and
\ref{fig:shift-orb} report the behavior of the Knight shift and of the orbital
shift, excluding the contribution of the macroscopic susceptibility.  We find
that the orbital susceptibility is the hardest to converge. This is coherent
with the fact that as a second order derivative of the total energy, it depends
on very fine details of the Fermi surface. On the other hand, the spin
susceptibility is obtained as the average
over the unit cell of the spin density. As such, it is comparatively insensitive
to details of the Fermi surface, and converges much faster with respect to the
smearing parameter. A similar hierarchy is obtained for the convergence
behavior of the Knight and orbital shifts (not including their respective
susceptibility). It should be noted that in the examples provided here, the
Knight shift is by far the largest component of the total NMR shifts.  Overall, we
expect the total NMR \textit{shielding} to be converged to better than $4\%$
with respect to smearing and $\vec{k}$-point density.  

On the other hand, convergence of the magnetic
susceptibility can prove quite arduous. For instance, the orbital
susceptibility of Aluminum varies from $-0.3$ to
$+5.6$\,$10^{-6}$\,cm$^3$\,mol$^{-1}$ within the temperature range
0.3\,eV to 0.1\,eV. Aluminum presents the slowest convergence of the three
metals studied in this work.



\subsection{Results and Discussion}

\subsubsection{Macroscopic Magnetic Susceptibility}
The computed magnetic susceptibility are referenced in
Tab.~\ref{tab:suscept}. Overall, agreement is very good. It 
contains a diamagnetic contribution from the core electrons. This contribution
is constant within the frozen core approximation and is computed once and for
all from an atomic code for each pseudo-potential. Tab.~\ref{tab:gas} compares
the spin and orbital susceptibilities of each metal to an electronic gas of
corresponding mean density. 

When examining the band structure of Aluminum, one finds that it is
quite similar to that of an homogeneous gas of equivalent density. As
a result, the non-interacting spin susceptibility and the Stoner
factor of these two systems are remarkably close. This indicates that
not only are their density of states at the Fermi level similar, but
also the Pauli-mediated behavior of the electrons with respect to a
pertubation of the spin population. On the other hand, the orbital
susceptibility of these two systems are quite different (note however
that for Aluminum, we did not achieve good convergence of this
quantity with respect to smearing). Indeed, in an ideal gas, the
contribution of lower lying electrons cancels-out exactly. Thus,
only electrons at the Fermi level contribute to the orbital
susceptibility. This is usually not true in more complex systems. Even small
differences between the band structures of Aluminum and the homogeneous gas
will result in appreciably different orbital
susceptibilities.

Lithium presents a case very different from the one above. Its
non-interacting spin susceptibility is much larger than that of the
homogeneous gas. As a result, the large polarization at the Fermi
level yields a large polarization of the lower-lying electronic
wavefunctions. The Stoner factor of Lithium is much larger than that of the
homogeneous electron gas. Interestingly, Lithium presents very little orbital
susceptibility.

Copper presents a different picture still. Indeed, it has a rather low
density of states at the Fermi level compared to the homogeneous gas.
As a result, both non-interacting and interacting spin-susceptibilities are
small. On the other hand, the large number of lower lying electrons, including
$d$ electrons, yield an
appreciable diamagnetic orbital susceptibility. As such, of the three
metals studied here, it is the only one with a diamagnetic
susceptibility. It is worthwhile to note that only the orbital
susceptibility can explain such a behavior, and that hence a complete
understanding of the susceptibility of Copper requires the computation of
both spin and orbital contributions.

\begin{table}[tbph]
  \begin{center}
     \begin{ruledtabular}
     \begin{tabular}{c|r@{\,$\pm$}lr@{\,$\pm$\!}lccc}
        Metal & \multicolumn{2}{c}{$\chi_s$} & \multicolumn{2}{c}{$\chi_o$}
              & $\chi_{core}$  & $\chi^{th}$ & Exp.{} \\ \hline
        Al &   17.7 & 0.2   &    1.9   &  5    & -3.0 &  16.6 & 16.5~[\onlinecite{carter:1977}] \\ 
        Li &   28.4 & 0.5   &    0.7   &  1    & -0.7 &  28.4 & 24.5$\pm0.3$~[\onlinecite{dugan:1997}] \\ 
        Cu &   10.8 & 0.2   &   -13.1  &  1    & -4.5 &  -6.8 &  -5.3~[\onlinecite{bowers:1956}] \\
     \end{tabular}
     \end{ruledtabular}
     \caption{Isotropic magnetic macroscopic susceptibility (in $10^{-6}$\,
              cm\,$^3$\,mol$^{-1}$, moles of atoms). The susceptibility contains three
              components: (i) the orbital susceptibility ($\chi_o$), (ii) the spin
              susceptibility ($\chi_s$), and (iii) the diamagnetic
              susceptibility of the core electrons ($\chi_{core}$). As
              shown in Fig.~\ref{fig:suscept-orb}, we were unable to
              converge the orbital susceptibility of Aluminum.   }
     \label{tab:suscept}
  \end{center}
\end{table}

\begin{table}[tbph]
  \begin{center}
    \begin{ruledtabular}
    \begin{tabular}{c|cccc}
      System & $\chi_s^0$ & Stoner & $\chi_s$ & $\chi_o$ \\\hline
      Al     &    13.2    &  1.34  &  17.7    &    1.9  \\
      gas    &    12.5    &  1.31  &  16.4    &   -4.2  \\\hline
      Li     &    15.5    &  1.83  &  28.4    &    0.7  \\
      gas    &    10.2    &  1.48  &  15.1    &   -3.4  \\\hline 
      Cu     &     9.5    &  1.14  &  10.8    &  -13.1  \\
      gas    &    15.3    &  1.18  &  18.1    &   -5.1  \\
    \end{tabular}
    \end{ruledtabular}
    \caption{Isotropic magnetic macroscopic susceptibilities (in $10^{-6}$\,
             cm\,$^3$\,mol$^{-1}$, moles of atoms). $\chi_s^0$ is the
             non-interacting spin susceptibility, $\chi_s$ the interacting spin
             susceptibility, and $\chi_o$ the orbital susceptibility. For comparison, the
             susceptibilities of a homogeneous gas of the same mean density as the system is
             given.}
    \label{tab:gas}
  \end{center}
\end{table}

\begin{figure}[htbp]
  \begin{center}
    \vspace{0.5cm}
    \includegraphics[width=0.85\columnwidth]{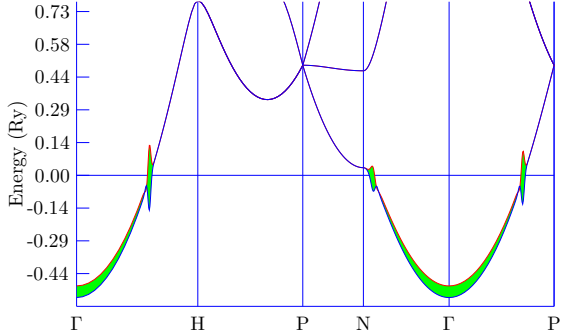}
    \caption{Band structure of lithium. The width of the
      line is representative of its contribution to the isotropic Knight shift. The
      Fermi energy is set to zero on the energy scale. Computations were done with a
      smearing of 0.2\,eV, for which the $\frac{1}{T}$ dependence is obvious at the
      Fermi energy. The ''divergence'' in $T$ disappears with the Brillouin zone
      integration.}
    \label{fig:bands}
  \end{center}
\end{figure}

\subsubsection{NMR shifts}
The computed isotropic NMR shifts are reported in
Tab.~\ref{tab:NMRshifts}. Tab.~\ref{tab:nonscfNMRshifts} also report
$\sigma_s/\sigma_s^0$, a quantity akin to the Stoner factor of the
susceptibility, where $\sigma_s$ is the Knight shift computed \emph{including}
self-consistency, and $\sigma_s^0$ the Knight shift computed \emph{without}
self-consistency.

The NMR shift of Aluminum results predominantly from the Knight shift.
It is worthwhile to note that the orbital and Knight \emph{shielding}
tensors are of similar magnitude, $-548$\,ppm and $1862$\,ppm
respectively. Yet, whereas the Knight contribution enters into the NMR
shift as a whole, the orbital part enters as a variation of the
absolute orbital shielding tensor between pure Al and ionic Al (which
presents no Knight shift), yielding a much smaller contribution.
Previous theoretical calculations\cite{mishra:1990} predict a Knight
shielding of $\sigma_s=1707$\,ppm. Although, the authors of
Ref.~\onlinecite{mishra:1990} do not compute NMR shifts comparable to
experiment, in the sense that they do not reference their results to a computed
zero-shift compound, their result is close to experimental value
because of the predominance of the Knight shift. As will be the case
for the other metals studied here, the ratio $\sigma_s/\sigma_s^0$ and
the Stoner factor are quite close in value. Indeed, both quantities
represent the same physical phenomena, namely the interplay between
the Kohn-Sham potential of the valence electrons and the spin
polarization at the Fermi level.

Again, the orbital shift of Lithium is by far smaller than its Knight shift. As
mentioned previously, the lower lying levels are heavily polarized
by electrons at the Fermi surface. The authors of
Ref.~\onlinecite{gaspari:1964} estimated the Knight shift of Lithium
including core-polarization. Even in this case, where from
Fig.~\ref{fig:bands} one would expect a rather high polarization, the
contribution is only of the order of 5\% of the whole (250\,ppm). More recently
Mishra \textit{et al} estimate a Knight shift of
301.9\,ppm\cite{mishra:1990}. Overall, our calculation agrees very
well with experimental values. The ratio $\sigma_s/\sigma_s^0$ is
relatively smaller than the Stoner factor. One should note that the
latter is a ratio of the average spin-polarization over the whole unit
cell, whereas the former is the ratio over the spin-polarization at a
single point of unit cell, namely the position of the Lithium nucleus.
The discrepancy between the two quantities implies simply that the
effect of the spin polarization is smaller at the nucleus than on
average across the cell.

Of the three metals studied here, Copper is the only one which
presents an appreciable orbital contribution to the NMR shift. It is
probably a result of the filled $d$-bands. Nonetheless, as
large as the orbital contribution may be, the Knight shift is larger
still. Interestingly, the computed absolute orbital shielding tensor
(including both valence and core contributions) is rather small
(26\,ppm). It would seem that a substantial paramagnetic
contribution from the valence electrons cancels out the substantial
diamagnetic contribution from the core electrons (computed to be
2171\,ppm). In other words, whereas in Li and Al, the reference
compound and the metals had similar orbital shielding tensor, the
orbital behavior of metallic Cu is very different from that of
Copper-Bromide. As was the case for the magnetic spin susceptibility,
the spin-polarization at the Fermi level has little effect on the
lower lying levels, resulting in a relatively small
$\sigma_s/\sigma_s^0$ ratio and Stoner factor.

\begin{table}[tbph]
  \begin{center}
     \begin{ruledtabular}
     \begin{tabular}{c|r@{\ $\pm$\negthickspace\negthickspace\negthickspace}lr@{\ $\pm$\negthickspace\negthickspace\negthickspace}lcc}
        Metal & \multicolumn{2}{c}{$\sigma_s$}
              & \multicolumn{2}{c}{$\sigma_o-\sigma_{ref}$}
              & $\delta$ 
              & Exp. \\ \hline
        Al & -1858 & 70  &  -16 & 8   & 1874 & 1640~[\onlinecite{sagalyn:1962}], \\
        Li &  -266 & 5   &  -15 & 1   &  281 &  260~[\onlinecite{carter:1977}] \\ 
        Cu & -2336 & 20  & -450 & 10  & 2786 & 2380~[\onlinecite{carter:1977}],\\
     \end{tabular}
     \end{ruledtabular}
     \caption{Isotropic NMR shifts of a few simple metals. For comparison, the
              orbital shielding with respect to the reference and the Knight shifts are given
              as well. The isotropic NMR shifts are given by the relationship $\delta 
              = -(\sigma_o+\sigma_s-\sigma_{ref})$. Estimates of the convergence with
              respect to temperature and Brillouin zone sampling are given in the first two
              columns.}
     \label{tab:NMRshifts}
  \end{center}
\end{table}

\begin{table}[tbph]
  \begin{center}
     \begin{ruledtabular}
       \begin{tabular}{c|cccccc}
       Metal & $\sigma_s^0$
             & $\sigma_s/\sigma_s^0$
             & $\sigma_s$
             & $\sigma_o-\sigma_{ref}$
             & $\delta$ 
             & Exp. \\ \hline
        Al & -1330 & 1.40 & -1858 &  -16 & 1874 & 1640~[\onlinecite{sagalyn:1962}], \\
        Li &  -157 & 1.69 &  -266 &  -15 &  281 &  260~[\onlinecite{carter:1977}] \\ 
        Cu & -2121 & 1.10 & -2336 & -604 & 2940 & 2380~[\onlinecite{carter:1977}],\\
     \end{tabular}
     \end{ruledtabular}
     \caption{Isotropic NMR shifts of a few simple metals.
     $\sigma_s^0$ is
     the \emph{non-interacting} Knight shift computed \emph{without} the
     self-consistent part of the pertubation. The ratio
     $\sigma_s/\sigma_s^0$ is
     the Knight shift equivalent of Stoner factor of the 
     spin susceptibility. Unsurprisingly, this ratio is quite close to the
     Stoner factor. Indeed both are a measure of the interplay between
     the spin-polarization at the Fermi level and lower-lying
     valence electrons.}
     \label{tab:nonscfNMRshifts}
  \end{center}
\end{table}

\section{Conclusions}
 We have presented a unified method for computing NMR shifts in
 metals. Our approach yields shifts which are directly comparable to
 experimental data, in the sense that both orbital and Knight shifts
 are computed. It was implemented within a pseudo-potential,
 plane-wave density functional theory code. All-electron quantities
 were recovered using the PAW approach. Gauge invariance was enforced
 with GIPAW. We compared results given by our approach to known
 analytical solutions for the homogeneous gas. Finally we successfully
 computed the NMR shifts of simple metals, with good comparison to
 experimental results. In conclusion, we have described a
 method which can accurately recover the NMR shifts of real metallic
 systems, thus
 allowing a better interpretation of NMR data. Next, we expect to study
 semi-metallic systems, such as graphite and nanotubes, for which an
 accurate description of both orbital and Knight shift is of paramount
 importance. 

\section*{Acknowledgment}
MA acknowledges support from MIT France and MURI grant DAAD 19-03-1-0169.

\appendix
\section{The Generalized $f$-sum rule}
\label{appendix}
Let $\mathcal{O}$ and $\mathcal{E}$ be odd and even operators respectively on time
reversal, i.e. for any real wave-functions $\ket{\Psi}$ and  $\ket{\Psi'}$:
\begin{equation}
  \bra{\Psi}\mathcal{O}\ket{\Psi'} = -\bra{\Psi'}\mathcal{O}\ket{\Psi}, \quad \bra{\Psi}\mathcal{E}\ket{\Psi'} = \bra{\Psi'}\mathcal{E}\ket{\Psi}
\end{equation}

Let $\ket{\Psi_i}$ be the eigen-wave-functions of the hamiltonian $\mathcal{H}$, with
eigenvalues $\epsilon_i$. Let $f(x)$ be a smearing function and $\sigma$ the
smearing. Then the occupation factors are defined as
$f_{j,i}=f(\frac{\epsilon_j - \epsilon_i}{\sigma})$ (where $i,j=F$ stands for
the Fermi energy $\epsilon_F$, and finally, let
\begin{gather}
	s = \sum_{i}\ \Re\left\{\bra{\Psi_i}\mathcal{O}\ \mathcal{G}(\epsilon_i)\
		\frac{1}{\imath}\left[\mathcal{E},\mathcal{H}\right]\ket{\Psi_i}\right\}\ \\
  \mathcal{G}(\epsilon_i) = \sum_{j} \frac{f_{F,j} - f_{F,i}}{\epsilon_j -\epsilon_i}\ket{\Psi_j}\bra{\Psi_j}
\end{gather}
where $\Re$ is the real part. Then, using the fact that
$\mathcal{H}\ket{\Psi_i}=\epsilon_i\ket{\Psi_i}$, we arrive at the expression:
\begin{equation}
	s = -\sum_{i,j}\left(f_{F,j}-f_{F,i}\right)\ 
		\Re\left[\bra{\Psi_i}\mathcal{O}\ket{\Psi_j}\bra{\Psi_j}
		\frac{1}{\imath}\mathcal{E}\ket{\Psi_i}\right]
\end{equation}
which can be separated into two sums:
\begin{multline}
	s = \sum_{i,j}f_{F,i}\Re\left[\bra{\Psi_i}\mathcal{O}\ket{\Psi_j}\bra{\Psi_j}
		\frac{1}{\imath}\mathcal{E}\ket{\Psi_i}\right] \\
    	    -\sum_{i,j}f_{F,j}\Re\left[\bra{\Psi_i}\mathcal{O}\ket{\Psi_j}\bra{\Psi_j}
		\frac{1}{\imath}\mathcal{E}\ket{\Psi_i}\right]
\end{multline}
Swapping dummy indexes in the second term:
\begin{multline}
	s = \sum_{i,j}f_{F,i}\Re\left[\bra{\Psi_i}\mathcal{O}\ket{\Psi_j}\bra{\Psi_j}
		\frac{1}{\imath}\mathcal{E}\ket{\Psi_i}\right] \\
    	    -\sum_{i,j}f_{F,\imath}\Re\left[\bra{\Psi_j}\mathcal{O}\ket{\Psi_j}\bra{\Psi_i}
		\frac{1}{\imath}\mathcal{E}\ket{\Psi_j}\right]
\end{multline}
Then, using the parity of $\mathcal{O}$ and $\mathcal{E}$:
\begin{multline}
	s = \sum_{i,j}f_{F,i}
		\Re\left[\bra{\Psi_i}\mathcal{O}\ket{\Psi_j}\bra{\Psi_j}
		\frac{1}{\imath}\mathcal{E}\ket{\Psi_i}\right] \\
    	    +\sum_{i,j}f_{F,i}
		\Re\left[\bra{\Psi_i}\mathcal{O}\ket{\Psi_j}\bra{\Psi_j}
		\frac{1}{\imath}\mathcal{E}\ket{\Psi_i}\right]
\end{multline}
After remarking that $\sum_j\ket{\Psi_j}\bra{\Psi_j}=1$:
\begin{equation}
	s = 2\sum_{i}f_{F,i}
		\Re\left[\bra{\Psi_i}\frac{1}{\imath}\mathcal{O}\mathcal{E}\ket{\Psi_i}\right]
\end{equation}
Expanding the real value, we arrive at the result:
\begin{multline}
  -\sum_i\ \Re\left\{\bra{\Psi_i}\mathcal{O}\ \mathcal{G}(\epsilon_i)\ 
		\frac{1}{\imath}\left[\mathcal{E},\mathcal{H}\right]\ket{\Psi_i}\right\}\ = \\
	 \sum_{i}f_{F,i}\bra{\Psi_i}\frac{1}{\imath}\left[\mathcal{E},\mathcal{O}\right]\ket{\Psi_i}\label{eq:sumrule}
\end{multline}
Expression \ref{eq:sumrule} and equation
(A7) in the appendix of Ref.~\onlinecite{pickard:2001} differ by the definition
of the Green function and the range of the sum over states. At zero
temperature and in insulators, the results are equivalent.

 \bibliography{biblio}

\begin{thebibliography}{33}
\expandafter\ifx\csname natexlab\endcsname\relax\def\natexlab#1{#1}\fi
\expandafter\ifx\csname bibnamefont\endcsname\relax
  \def\bibnamefont#1{#1}\fi
\expandafter\ifx\csname bibfnamefont\endcsname\relax
  \def\bibfnamefont#1{#1}\fi
\expandafter\ifx\csname citenamefont\endcsname\relax
  \def\citenamefont#1{#1}\fi
\expandafter\ifx\csname url\endcsname\relax
  \def\url#1{\texttt{#1}}\fi
\expandafter\ifx\csname urlprefix\endcsname\relax\def\urlprefix{URL }\fi
\providecommand{\bibinfo}[2]{#2}
\providecommand{\eprint}[2][]{\url{#2}}

\bibitem[{\citenamefont{Grant and Harris}(1996)}]{grant}
\bibinfo{editor}{\bibfnamefont{D.~M.} \bibnamefont{Grant}} \bibnamefont{and}
  \bibinfo{editor}{\bibfnamefont{R.~K.} \bibnamefont{Harris}}, eds.,
  \emph{\bibinfo{title}{The Encyclopedia of {NMR}}}
  (\bibinfo{publisher}{Wiley}, \bibinfo{address}{London},
  \bibinfo{year}{1996}).

\bibitem[{\citenamefont{Pines}(1997)}]{pines:1997}
\bibinfo{author}{\bibfnamefont{D.}~\bibnamefont{Pines}}, \bibinfo{journal}{Z.\
  Phys.~B} \textbf{\bibinfo{volume}{103}}, \bibinfo{pages}{129}
  (\bibinfo{year}{1997}).

\bibitem[{\citenamefont{Keith and Bader}(1992)}]{keith:1992}
\bibinfo{author}{\bibfnamefont{T.~A.} \bibnamefont{Keith}} \bibnamefont{and}
  \bibinfo{author}{\bibfnamefont{R.~F.~W.} \bibnamefont{Bader}},
  \bibinfo{journal}{Chem.\ Phys.\ Lett.} \textbf{\bibinfo{volume}{194}},
  \bibinfo{pages}{1} (\bibinfo{year}{1992}).

\bibitem[{\citenamefont{Pickard and Mauri}(2001)}]{pickard:2001}
\bibinfo{author}{\bibfnamefont{C.~J.} \bibnamefont{Pickard}} \bibnamefont{and}
  \bibinfo{author}{\bibfnamefont{F.}~\bibnamefont{Mauri}},
  \bibinfo{journal}{Phys.\ Rev.\ B:\ Condens.\ Matter}
  \textbf{\bibinfo{volume}{63}}, \bibinfo{pages}{245101}
  (\bibinfo{year}{2001}).

\bibitem[{\citenamefont{Sebastiani and Parrinello}(2001)}]{sebastiani:2001}
\bibinfo{author}{\bibfnamefont{D.}~\bibnamefont{Sebastiani}} \bibnamefont{and}
  \bibinfo{author}{\bibfnamefont{M.}~\bibnamefont{Parrinello}},
  \bibinfo{journal}{J.~Phys.\ Chem.~A} \textbf{\bibinfo{volume}{105}},
  \bibinfo{pages}{1951} (\bibinfo{year}{2001}).

\bibitem[{\citenamefont{Profeta et~al.}(2003)\citenamefont{Profeta, Mauri, and
  Pickard}}]{profeta:2003}
\bibinfo{author}{\bibfnamefont{M.}~\bibnamefont{Profeta}},
  \bibinfo{author}{\bibfnamefont{F.}~\bibnamefont{Mauri}}, \bibnamefont{and}
  \bibinfo{author}{\bibfnamefont{C.~J.} \bibnamefont{Pickard}},
  \bibinfo{journal}{J.~Am.\ Chem.\ Soc.} \textbf{\bibinfo{volume}{125}},
  \bibinfo{pages}{541} (\bibinfo{year}{2003}).

\bibitem[{\citenamefont{Umari and Pasquarello}(2005)}]{umari:2005}
\bibinfo{author}{\bibfnamefont{P.}~\bibnamefont{Umari}} \bibnamefont{and}
  \bibinfo{author}{\bibfnamefont{A.}~\bibnamefont{Pasquarello}},
  \bibinfo{journal}{Phys.\ Rev.\ Lett.} \textbf{\bibinfo{volume}{95}},
  \bibinfo{pages}{137401} (\bibinfo{year}{2005}).

\bibitem[{\citenamefont{Tripathi et~al.}(1981)\citenamefont{Tripathi, Das, ,
  Misra, and Mahanti}}]{tripathi:1981}
\bibinfo{author}{\bibfnamefont{G.~S.} \bibnamefont{Tripathi}},
  \bibinfo{author}{\bibfnamefont{L.~K.} \bibnamefont{Das}}, ,
  \bibinfo{author}{\bibfnamefont{P.~K.} \bibnamefont{Misra}}, \bibnamefont{and}
  \bibinfo{author}{\bibfnamefont{S.~D.} \bibnamefont{Mahanti}},
  \bibinfo{journal}{Solid State Commun.} \textbf{\bibinfo{volume}{38}},
  \bibinfo{pages}{1207} (\bibinfo{year}{1981}).

\bibitem[{\citenamefont{Pavarini and Mazin}(2001)}]{pavarini:2001}
\bibinfo{author}{\bibfnamefont{E.}~\bibnamefont{Pavarini}} \bibnamefont{and}
  \bibinfo{author}{\bibfnamefont{I.~I.} \bibnamefont{Mazin}},
  \bibinfo{journal}{Phys.\ Rev.\ B:\ Condens.\ Matter}
  \textbf{\bibinfo{volume}{64}}, \bibinfo{pages}{140504}
  (\bibinfo{year}{2001}).

\bibitem[{\citenamefont{Lauginie et~al.}(1988)\citenamefont{Lauginie,
  {Estrade-Szwarckopf}, Rousseau, and Conard}}]{lauginie:1988}
\bibinfo{author}{\bibfnamefont{P.}~\bibnamefont{Lauginie}},
  \bibinfo{author}{\bibfnamefont{H.}~\bibnamefont{{Estrade-Szwarckopf}}},
  \bibinfo{author}{\bibfnamefont{B.}~\bibnamefont{Rousseau}}, \bibnamefont{and}
  \bibinfo{author}{\bibfnamefont{J.}~\bibnamefont{Conard}},
  \bibinfo{journal}{{CR.\ Acad.\ Sci.\ Paris}}
  \textbf{\bibinfo{volume}{307II}}, \bibinfo{pages}{1693}
  (\bibinfo{year}{1988}).

\bibitem[{\citenamefont{Hiroyama and Kume}(1988)}]{hyroyama:1988}
\bibinfo{author}{\bibfnamefont{Y.}~\bibnamefont{Hiroyama}} \bibnamefont{and}
  \bibinfo{author}{\bibfnamefont{K.}~\bibnamefont{Kume}},
  \bibinfo{journal}{Solid State Commun.} \textbf{\bibinfo{volume}{65}},
  \bibinfo{pages}{617} (\bibinfo{year}{1988}).

\bibitem[{\citenamefont{Lauginie et~al.}(1993)\citenamefont{Lauginie,
  Messaoudi, and Conard}}]{lauginie:1993}
\bibinfo{author}{\bibfnamefont{P.}~\bibnamefont{Lauginie}},
  \bibinfo{author}{\bibfnamefont{A.}~\bibnamefont{Messaoudi}},
  \bibnamefont{and} \bibinfo{author}{\bibfnamefont{J.}~\bibnamefont{Conard}},
  \bibinfo{journal}{Synthetic Metals} \textbf{\bibinfo{volume}{56}},
  \bibinfo{pages}{3002} (\bibinfo{year}{1993}).

\bibitem[{\citenamefont{Kobayashi and Tsukada}(1988)}]{kobayashi:1988}
\bibinfo{author}{\bibfnamefont{K.}~\bibnamefont{Kobayashi}} \bibnamefont{and}
  \bibinfo{author}{\bibfnamefont{M.}~\bibnamefont{Tsukada}},
  \bibinfo{journal}{Phys.\ Rev.\ B:\ Condens.\ Matter}
  \textbf{\bibinfo{volume}{38}}, \bibinfo{pages}{8566} (\bibinfo{year}{1988}).

\bibitem[{\citenamefont{Ajiki and Ando}(1995)}]{ajiki:1995}
\bibinfo{author}{\bibfnamefont{H.}~\bibnamefont{Ajiki}} \bibnamefont{and}
  \bibinfo{author}{\bibfnamefont{T.}~\bibnamefont{Ando}},
  \bibinfo{journal}{{J.~Phys.\ Soc.\ Japan}} \textbf{\bibinfo{volume}{64}},
  \bibinfo{pages}{4382} (\bibinfo{year}{1995}).

\bibitem[{\citenamefont{Stokes et~al.}(1982{\natexlab{a}})\citenamefont{Stokes,
  Rhodes, Wang, Slichter, and Sinfelt}}]{stokes:1982-i}
\bibinfo{author}{\bibfnamefont{H.~T.} \bibnamefont{Stokes}},
  \bibinfo{author}{\bibfnamefont{H.~E.} \bibnamefont{Rhodes}},
  \bibinfo{author}{\bibfnamefont{P.-K.} \bibnamefont{Wang}},
  \bibinfo{author}{\bibfnamefont{C.~P.} \bibnamefont{Slichter}},
  \bibnamefont{and} \bibinfo{author}{\bibfnamefont{J.~H.}
  \bibnamefont{Sinfelt}}, \bibinfo{journal}{Phys.\ Rev.\ B:\ Condens.\ Matter}
  \textbf{\bibinfo{volume}{26}}, \bibinfo{pages}{3559}
  (\bibinfo{year}{1982}{\natexlab{a}}).

\bibitem[{\citenamefont{Stokes et~al.}(1982{\natexlab{b}})\citenamefont{Stokes,
  Rhodes, Wang, Slichter, and Sinfelt}}]{stokes:1982-iii}
\bibinfo{author}{\bibfnamefont{H.~T.} \bibnamefont{Stokes}},
  \bibinfo{author}{\bibfnamefont{H.~E.} \bibnamefont{Rhodes}},
  \bibinfo{author}{\bibfnamefont{P.-K.} \bibnamefont{Wang}},
  \bibinfo{author}{\bibfnamefont{C.~P.} \bibnamefont{Slichter}},
  \bibnamefont{and} \bibinfo{author}{\bibfnamefont{J.~H.}
  \bibnamefont{Sinfelt}}, \bibinfo{journal}{Phys.\ Rev.\ B:\ Condens.\ Matter}
  \textbf{\bibinfo{volume}{26}}, \bibinfo{pages}{3575}
  (\bibinfo{year}{1982}{\natexlab{b}}).

\bibitem[{\citenamefont{Makowka et~al.}(1985)\citenamefont{Makowka, Slichter,
  and Sinfelt}}]{makowka:1985}
\bibinfo{author}{\bibfnamefont{C.~D.} \bibnamefont{Makowka}},
  \bibinfo{author}{\bibfnamefont{C.~P.} \bibnamefont{Slichter}},
  \bibnamefont{and} \bibinfo{author}{\bibfnamefont{J.~H.}
  \bibnamefont{Sinfelt}}, \bibinfo{journal}{Phys.\ Rev.\ B:\ Condens.\ Matter}
  \textbf{\bibinfo{volume}{31}}, \bibinfo{pages}{5663} (\bibinfo{year}{1985}).

\bibitem[{\citenamefont{Vuissoz et~al.}(1999)\citenamefont{Vuissoz, Ansermet,
  and Wieckowski}}]{vuissoz:1999}
\bibinfo{author}{\bibfnamefont{P.-A.} \bibnamefont{Vuissoz}},
  \bibinfo{author}{\bibfnamefont{J.-P.} \bibnamefont{Ansermet}},
  \bibnamefont{and}
  \bibinfo{author}{\bibfnamefont{A.}~\bibnamefont{Wieckowski}},
  \bibinfo{journal}{Phys.\ Rev.\ B:\ Condens.\ Matter}
  \textbf{\bibinfo{volume}{83}}, \bibinfo{pages}{2457} (\bibinfo{year}{1999}).

\bibitem[{\citenamefont{Bl{\"o}chl}(1994)}]{blochl:1994}
\bibinfo{author}{\bibfnamefont{P.~E.} \bibnamefont{Bl{\"o}chl}},
  \bibinfo{journal}{Phys.\ Rev.\ B:\ Condens.\ Matter}
  \textbf{\bibinfo{volume}{50}}, \bibinfo{pages}{17953} (\bibinfo{year}{1994}).

\bibitem[{\citenamefont{de~Gironcoli}(1995)}]{gironcoli:1995}
\bibinfo{author}{\bibfnamefont{S.}~\bibnamefont{de~Gironcoli}},
  \bibinfo{journal}{Phys.\ Rev.\ B:\ Condens.\ Matter}
  \textbf{\bibinfo{volume}{51}}, \bibinfo{pages}{6773} (\bibinfo{year}{1995}).

\bibitem[{\citenamefont{van~de Walle and Bl{\"o}chl}(1993)}]{blochl:1993}
\bibinfo{author}{\bibfnamefont{C.~G.} \bibnamefont{van~de Walle}}
  \bibnamefont{and} \bibinfo{author}{\bibfnamefont{P.~E.}
  \bibnamefont{Bl{\"o}chl}}, \bibinfo{journal}{Phys.\ Rev.\ B:\ Condens.\
  Matter} \textbf{\bibinfo{volume}{47}}, \bibinfo{pages}{4244}
  (\bibinfo{year}{1993}).

\bibitem[{\citenamefont{Pacchioni et~al.}(2000)\citenamefont{Pacchioni,
  Frigoli, Ricci, and Weil}}]{pacchioni:2000}
\bibinfo{author}{\bibfnamefont{G.}~\bibnamefont{Pacchioni}},
  \bibinfo{author}{\bibfnamefont{F.}~\bibnamefont{Frigoli}},
  \bibinfo{author}{\bibfnamefont{D.}~\bibnamefont{Ricci}}, \bibnamefont{and}
  \bibinfo{author}{\bibfnamefont{J.~A.} \bibnamefont{Weil}},
  \bibinfo{journal}{Phys.\ Rev.\ B:\ Condens.\ Matter}
  \textbf{\bibinfo{volume}{63}}, \bibinfo{pages}{54102} (\bibinfo{year}{2000}).

\bibitem[{\citenamefont{Ashcroft and Mermin}(1976)}]{ashcroft}
\bibinfo{author}{\bibfnamefont{N.~W.} \bibnamefont{Ashcroft}} \bibnamefont{and}
  \bibinfo{author}{\bibfnamefont{N.~D.} \bibnamefont{Mermin}},
  \emph{\bibinfo{title}{Solid State Physics}}
  (\bibinfo{publisher}{Brooks/Cole}, \bibinfo{year}{1976}).

\bibitem[{\citenamefont{Perdew et~al.}(1996)\citenamefont{Perdew, Burke, and
  Ernzerhof}}]{perdew:1996}
\bibinfo{author}{\bibfnamefont{J.~P.} \bibnamefont{Perdew}},
  \bibinfo{author}{\bibfnamefont{K.}~\bibnamefont{Burke}}, \bibnamefont{and}
  \bibinfo{author}{\bibfnamefont{M.}~\bibnamefont{Ernzerhof}},
  \bibinfo{journal}{Phys.\ Rev.\ Lett.} \textbf{\bibinfo{volume}{77}},
  \bibinfo{pages}{3865} (\bibinfo{year}{1996}).

\bibitem[{\citenamefont{Marzari et~al.}(1999)\citenamefont{Marzari, Vanderbilt,
  {De Vita}, and Payne}}]{marzari:1999}
\bibinfo{author}{\bibfnamefont{N.}~\bibnamefont{Marzari}},
  \bibinfo{author}{\bibfnamefont{D.}~\bibnamefont{Vanderbilt}},
  \bibinfo{author}{\bibfnamefont{A.}~\bibnamefont{{De Vita}}},
  \bibnamefont{and} \bibinfo{author}{\bibfnamefont{M.~C.} \bibnamefont{Payne}},
  \bibinfo{journal}{Phys.\ Rev.\ Lett.} \textbf{\bibinfo{volume}{82}},
  \bibinfo{pages}{3296} (\bibinfo{year}{1999}).

\bibitem[{\citenamefont{Troullier and Martins}(1991)}]{troullier:1991}
\bibinfo{author}{\bibfnamefont{N.}~\bibnamefont{Troullier}} \bibnamefont{and}
  \bibinfo{author}{\bibfnamefont{J.~L.} \bibnamefont{Martins}},
  \bibinfo{journal}{Phys.\ Rev.\ B:\ Condens.\ Matter}
  \textbf{\bibinfo{volume}{43}}, \bibinfo{pages}{1993} (\bibinfo{year}{1991}).

\bibitem[{\citenamefont{Monkhorst and Pack}(1976)}]{monkhorst:1976}
\bibinfo{author}{\bibfnamefont{H.~J.} \bibnamefont{Monkhorst}}
  \bibnamefont{and} \bibinfo{author}{\bibfnamefont{J.~D.} \bibnamefont{Pack}},
  \bibinfo{journal}{Phys.\ Rev.\ B:\ Condens.\ Matter}
  \textbf{\bibinfo{volume}{13}}, \bibinfo{pages}{5188} (\bibinfo{year}{1976}).

\bibitem[{\citenamefont{Carter et~al.}(1977)\citenamefont{Carter, Benett, and
  Kahan}}]{carter:1977}
\bibinfo{author}{\bibfnamefont{G.~C.} \bibnamefont{Carter}},
  \bibinfo{author}{\bibfnamefont{L.~H.} \bibnamefont{Benett}},
  \bibnamefont{and} \bibinfo{author}{\bibfnamefont{D.~J.} \bibnamefont{Kahan}},
  \bibinfo{journal}{Progress in Material Science}
  \textbf{\bibinfo{volume}{20}}, \bibinfo{pages}{1} (\bibinfo{year}{1977}).

\bibitem[{\citenamefont{Dugan}(1997)}]{dugan:1997}
\bibinfo{author}{\bibfnamefont{D.}~\bibnamefont{Dugan}},
  \bibinfo{journal}{Phys.\ Rev.\ B:\ Condens.\ Matter}
  \textbf{\bibinfo{volume}{57}}, \bibinfo{pages}{7759} (\bibinfo{year}{1997}).

\bibitem[{\citenamefont{Bowers}(1956)}]{bowers:1956}
\bibinfo{author}{\bibfnamefont{R.}~\bibnamefont{Bowers}},
  \bibinfo{journal}{Phys.\ Rev.} \textbf{\bibinfo{volume}{102}},
  \bibinfo{pages}{1486} (\bibinfo{year}{1956}).

\bibitem[{\citenamefont{Mishra et~al.}(1990)\citenamefont{Mishra, Das, Sahu,
  Tripathi, and Misra}}]{mishra:1990}
\bibinfo{author}{\bibfnamefont{B.}~\bibnamefont{Mishra}},
  \bibinfo{author}{\bibfnamefont{L.~K.} \bibnamefont{Das}},
  \bibinfo{author}{\bibfnamefont{T.}~\bibnamefont{Sahu}},
  \bibinfo{author}{\bibfnamefont{G.~S.} \bibnamefont{Tripathi}},
  \bibnamefont{and} \bibinfo{author}{\bibfnamefont{P.~K.} \bibnamefont{Misra}},
  \bibinfo{journal}{J~Phys.\ : Cond.\ Mat.} \textbf{\bibinfo{volume}{2}},
  \bibinfo{pages}{9891} (\bibinfo{year}{1990}).

\bibitem[{\citenamefont{Gaspari et~al.}(1964)\citenamefont{Gaspari, Shyu, and
  Das}}]{gaspari:1964}
\bibinfo{author}{\bibfnamefont{G.~D.} \bibnamefont{Gaspari}},
  \bibinfo{author}{\bibfnamefont{W.}~\bibnamefont{Shyu}}, \bibnamefont{and}
  \bibinfo{author}{\bibfnamefont{T.~P.} \bibnamefont{Das}},
  \bibinfo{journal}{Phys.\ Rev.} \textbf{\bibinfo{volume}{134}},
  \bibinfo{pages}{A852} (\bibinfo{year}{1964}).

\bibitem[{\citenamefont{Sagalyn and Hofmann}(1962)}]{sagalyn:1962}
\bibinfo{author}{\bibfnamefont{P.~L.} \bibnamefont{Sagalyn}} \bibnamefont{and}
  \bibinfo{author}{\bibfnamefont{J.~A.} \bibnamefont{Hofmann}},
  \bibinfo{journal}{Phys.\ Rev.} \textbf{\bibinfo{volume}{127}},
  \bibinfo{pages}{68} (\bibinfo{year}{1962}).

\end{thebibliography}
\end{document}